\documentclass[12pt,preprint]{aastex}

\def\Epk{E_{\rm pk}}

\def\F0{F_{\rm 0}}

\def\t0{t_{\rm 0}}
\def\Fm{F_{\rm m}}
\def\tm{t_{\rm m}}
\def\Em{E_{\rm m}}
\def\taum{\tau_{\rm m}}

\def\P0{\Phi_{\rm 0}}
\def\tP0{\tilde{\Phi_{\rm 0}}}
\def\E00{E_{\rm pk,0}}
\def\td{t_{\rm d}}
\def\tr{t_{\rm r}}
\def\N0{N_{\rm 0}}
\def\F0{F_{\rm 0}}

\def\tr{\tau_{\rm r}}

\def\td{\tau_{\rm d}}

\def\tr{\tau_{\rm r}}
\def\td{\tau_{\rm d}}
\def\trd{\tau_{\rm r,d}}
\newcommand{\ltsima} {$\; \buildrel < \over \sim \;$}
\newcommand{\gtsima} {$\; \buildrel > \over \sim \;$}
\newcommand{\lta} {\lower.5ex\hbox{\ltsima}}
\newcommand{\gta} {\lower.5ex\hbox{\gtsima}}

\slugcomment{Date: \today ; Submitted to the ApJ}

\begin{document}

\title{Search for Relativistic Curvature Effects in Gamma-Ray Burst Pulses}

\author{Dan Kocevski\altaffilmark{1}, Felix Ryde \altaffilmark{1,2,3}, and
Edison Liang\altaffilmark{1}}

\altaffiltext{1}{Department of Physics and Astronomy, Rice
University, Houston, TX 77005}

\altaffiltext{2}{Center for Space Science and Astrophysics,
Stanford University, Stanford, CA 94305}

\altaffiltext{3}{Present address: Stockholm Observatory, AlbaNova,
SE-106 91 Stockholm, Sweden}

\email{kocevski@rice.edu}

\email{felix@ahoor.stanford.edu}

\email{liang@spacsun.rice.edu}

\begin{abstract}
We analyze the time profiles of individual gamma-ray burst (GRB)
pulses, that are longer than 2 s, by modelling them with
analytical functions that are based on physical first principles
and well-established empirical descriptions of GRB spectral
evolution. These analytical profiles are independent of the
emission mechanism and can be used to model both the rise and
decay profiles allowing for the study of the entire pulse
light-curve. Using this method, we have studied a sample of 77
individual GRB pulses, allowing us to examine the fluence, pulse
width, asymmetry, and rise and decay power-law distributions. We
find that the rise phase is best modelled with a power law of
average index $r = 1.31 \pm 0.11$ and that the average decay phase
has an index of $d =  2.39 \pm 0.12$. We also find that the ratio
between the rise and decay times (the pulse asymmetry) exhibited
by the GRB pulse shape has an average value of 0.47 which varies
little from pulse to pulse and is independent of pulse duration or
intensity.  Although this asymmetry is largely uncorrelated to
other pulse properties, a statistically significant trend is
observed between the pulse asymmetry and the decay power law
index, possibly hinting at the underlying physics. We compare
these parameters with those predicted to occur if individual pulse
shapes are created purely by relativistic curvature effects in the
context of the fireball model, a process that makes specific
predictions about the shape of GRB pulses. The decay index
distribution obtained from our sample shows that the average GRB
pulse fades faster than the value predicted by curvature effects,
with only 39$\%$ of our sample being consistent with the curvature
model. We discuss several refinements of the relativistic
curvature scenario that could naturally account for these observed
deviations, such as symmetry breaking and varying relative
time-scales within individual pulses.
\end{abstract}


\section*{Introduction}

The temporal structure of gamma-ray bursts (GRBs) varies
drastically, with no apparent pattern among bursts.  Of the
$>$2704 GRBs detected by the BATSE instrument onboard the Compton
Gamma Ray Observatory (CGRO) \citep{fish94}, less than 10 $\%$ of
the detected light-curve profiles can be categorized as being
similar in overall shape or morphology, in that they comprise
coherent structures, or pulses of radiation. It is generally
believed that these pulses represent the fundamental constituent
of GRB time profiles (light curves), and appear as asymmetric
pulses with a fast rise and a slower decay, often denoted FRED
"Fast Rise and Exponential Decay" (see Figure 1). Each pulse is
assumed to be associated with a separate emission episode, with
complex GRBs being superpositions of several such episodes, of
varying amplitude and intensity. Although the duration and
amplitude of the FRED pulses vary considerably, the shape is the
only recurring pattern that can be distinguished among the vast
range of complex GRB light curves that have been observed.

The observed $\gamma$-ray pulses are believed to be produced in a
highly relativistic outflow, an expanding and collimated fireball,
based on the large energies and the short time scales involved.
There are several possible mechanisms underlying the emission,
although the commonly assumed scenario is that individual pulses
are created, when shocks internal to the relativistic outflow,
drain the kinetic energy and accelerate leptons which radiate. In
this paper we base our discussion on this standard, fireball
model, in which the $\gamma$-rays arise from the internal shocks
at a distance of $R \sim 10^{13}-10^{17}$ cm from the initial
source \citep{piran99}.  In the context of the fireball model, the
episodic nature of the outflow causes inhomogeneities in the wind
to collide and thus create the shocks.  The dominant emission
mechanisms are most probably non-thermal synchrotron
\citep{tavani, LP01}) and/or inverse Compton emission
\citep{LKB03}, but there have been other suggestions, for
instance, thermal, saturated Comptonization \citep{liang}. The
simplest scenario here is to assume an impulsive heating of the
leptons and a subsequent cooling and emission.  Therefore, the
rise phase of the pulse is attributed to the energization of the
shell and the decay phase reflects the cooling of the energized
particles.

As many authors have pointed out, this cooling interpretation does
not work well when applied to non-thermal synchrotron and/or
inverse Compton emission because the radiative cooling timescales
alone are typically much too short to explain the pulse durations.
For example, if we interpret the average break energy of a GRB
spectrum ($\sim E_{pk}$) as the characteristic energy of
synchrotron self-absorption, then the resulting magnetic field
must be extremely high, about $1\times10^{7}$ to $1\times10^{8}$
Gauss. Similar estimates are obtained if we assume equipartition
conditions between the lepton energy and the magnetic field.  In
either case, such a high $B$ field would create a synchrotron
cooling timescale of the order of $10^{-5}$ seconds in the
comoving frame \citep{Wu}. One resolution to this problem is to
introduce relativistic effects. If the relativistic fireball
expands with a Lorentz factor of $\Gamma$ $>$ 100, then the
geometry of the shell would make radiation emitted off the line of
sight delayed and affected by a varying Doppler boost (see Figure
2). This would cause the GRB spectra to evolve to lower energies
and produce decay profiles much longer than the microscopic
cooling timescale. These relativistic curvature effects would
produce a signature decay profile that can be obtained
analytically and searched for in a sample of GRB pulses. Thus it
is of great importance to characterize the individual pulse
profiles within GRBs and study their parameter distributions.

Several investigations along the lines of pulse modelling have
perviously been made.  Initially, the pulse profiles were modelled
by "stretched exponential" functions, both for the rise phase and
for the decay phase \citep{nor96,lee1, lee2}:
\begin{equation}
F(t) = \F0 e^{ -(\vert t-t_{\rm max}\vert
/\trd)^{\nu}},\label{stretch}
\end{equation}
where $t_{\rm max}$ is the time of the maximum flux, $\F0$, of the
pulse, $\trd$ are the time constants for the rise and the decay
phases, respectively, and $\nu$ is the peakedness
parameter\footnote{For $\nu > 1$, equation (\ref{stretch}) is,
strictly speaking, a compressed exponential.}. Such a function is
very flexible which makes it possible to describe the whole shape
of most pulses, and to quantify the  characteristics of the pulses
for a statistical analysis, most notably their location,
amplitude, width, rise phase, transition phase, and decay phase.
\citet{nor96} studied a sample of bursts observed by BATSE and
found that the decay generally lies between a pure exponential
($\nu = 1$) and a Gaussian ($ \nu = 2$). \citet{lee1, lee2}
studied approximately 2500 pulse structures, in individual energy
channels, using the high time resolution BATSE TTS data type. They
confirmed the general behavior, namely, that pulses tend to have
shorter rise times than decay times.

The stretched exponential was introduced because of its very
flexible nature, although recently \citet{RS00, RS02} have
proposed that the decay phase would be better described by a power
law, which in terms of energy flux is :
\begin{equation}
F(t) =   \F0 \left(1+\frac{t}{T(d-1)} \right)^{-d},\label{FE}
\end{equation}
See \S \ref{sec:decay} for the derivation of equation (\ref{FE})
and definition of $d$.  The motivation for this type of shape is
entirely based on empirical relations describing the evolution of
the GRB spectra during the decay phase of individual pulses.

\citet{RP} have recently shown that the form expressed in equation
(\ref{FE}) can be produced through simple relativistic kinematics
when applied to a spherical shell expanding at extreme
relativistic velocity. The curvature of a relativistic fireball
would make the photons emitted off the line of sight (LOS) delayed
and affected by a varying Doppler boost due to the increasing
angle at which the photons were emitted with respect to the
observer (See Figure 2 for an illustration).  They show that this
Doppler boosting of off-axis photons can reproduce the two well
known empirical correlations observed in the GRB spectra, namely
the hardness-intensity and hardness-fluence correlations (HIC and
HFC, respectively), in a manner that is highly predictive of the
resulting pulse profile.\footnote{ See \citet{krm01} for the
effects on the morphology of a complex GRB light curve.} Following
\citet{RP}, we can derive the expected emission profile from a
spherical shell that radiates for an infinitesimal period of time
at a peak energy $E'_{pk}$ in the comoving frame.  The Lorentz
boosting factor for transformations from the comoving frame to the
observers frame of photons emitted from different locations on the
surface of a spherical shell is given by
\begin{equation} \label{doppler}
{\cal D} (\Gamma,\mu)=\frac{1}{\Gamma (1-\beta \mu)} =\Gamma
(1+\beta \mu'), \label{boost}
\end{equation}
where the angle $\theta \equiv \arccos \mu$ shown in figure 2. If
we define the point where the flow velocity is parallel to the
line of sight (LOS) to be $\theta=0$, then the difference in light
travel time between photons emitted along the LOS and photons
emitted at an angle $\theta$ is given by $\Delta t =
R_0(1-\mu)/c$, which gives $\mu= 1-c\Delta t/R_0$. Substituting
this into the Lorentz boost factor, we find
\begin{equation}
{\cal D} (\Gamma,\Delta t)=\frac{1}{\Gamma (1-\beta + \beta c
\Delta t/R_0)}.
\end{equation}
For extremely relativistic outflows $\beta = 1-(2\Gamma^2)^{-1}$,
and the boost factor becomes
\begin{equation} {\cal D} (\Gamma,\Delta
t)= 2 \Gamma (1+\Delta t/\tau_{\rm ang})^{-1}, \,\, \, \tau_{\rm
ang}\equiv R_0/(2\Gamma^2 c). \label{Dt}
\end{equation}
The outcome of such a Doppler profile is that if the emitted
spectra from different parts of the shell are identical, then the
observed spectra will be gradually redshifted in time by a factor
of $\cal D$ as photons from different parts of the shell are
received by the observer. The peak energy of the spectra will be
observed to evolve as
\begin{equation}
\Epk(t)=E'_{\rm pk}{\cal D}=\frac{\E00}{(1+t/\tau_{\rm ang})}
\end{equation}
Where $E'_{\rm pk}$ is the peak energy in the comoving frame. By
similar arguments, \citet{RP} show that the resulting bolometric
energy flux should evolve as
\begin{equation}
F_{\rm bol}(t) = F_{0} {\cal D}^{2} = \frac{F_{0}}{(1+t/\tau_{\rm
ang})^2}
 \label{Ft_rel}
\end{equation}
Therefore, the decay of the resulting light curve, produced purely
by relativistic effects, should exhibit a distinct profile, given
by equation (\ref{Ft_rel}). This result becomes more complicated
when one considers the three primary timescales relevant for the
shape of the pulse: the shell crossing time, the angular spreading
time, and the cooling time. The first two timescales are
comparable and dominate over the third and compete with each other
to produce the observed results. The authors find that for the
case of "pure" spherical curvature, where the curvature timescale
dominates over the intrinsic light-curve profile (the shell
crossing time), the convolved light curves indeed reach the
behavior of equation (\ref{Ft_rel}) fairly quickly. However, if
the crossing timescale and the curvature timescale are comparable,
then the resulting light curve will have some convolved shape and
the index value of 2 will only be manifested at late times in the
pulse decay.

The \citet{RP} treatment assumes that the variations in the low-
and high-energy power law indices ($\alpha$ and $\beta$) that are
commonly used to describe the GRB spectra tend to be small over
the length of the pulse, which may not necessarily be the case. In
order to account for this, \citet{Fen96} and \citet{FenRik} used a
photon number spectrum which they described by a single power law
index equal to the average value of $(\alpha+\beta)/2$. They
examined the resulting light curve and spectral evolution that
would be expected if the FRED profile were due primarily to
relativistic effects.  They conclude that the decay phase of FRED
profiles should be independent of intrinsic spectral variations
and should scale roughly as $T^{-\alpha-1}$.

Therefore, one of the primary goals of this paper is to model the
FRED light curves using an analytic function derived from physical
first principles and relations describing the spectral evolution
of GRBs to obtain parameters that uniquely quantify the shape of
smooth pulses and compare those values to the predicted signatures
of relativistic kinematics.  Preliminary results of this analysis
have been presented in \citet{KL2} and \citet{RKL}, where we
derived analytic models and tested for a small number of pulses.
This paper expands on these previous results to include a much
larger sample and a more detailed analysis of the distribution of
pulse attributes. In $\S 1$, we review the empirical
hardness-intensity and hardness-fluence correlations which are
crucial to the derivation of an analytic pulse model. In $\S 2$,
we derive an analytic function based on the above-mentioned
spectral correlations. In $\S 3$, we define a sample of FRED
pulses observed by CGRO to which we fit our function to obtain a
distribution of model parameters, most notably the decay power law
index and pulse asymmetries.  In $\S 4$, we discuss our measured
parameter distributions and compare them to the values predicted
by relativistic geometry.  We discuss several mechanisms that can
skew the distributions away from the expected values.

\section{Spectral Correlations} \label{sec:spectral}

Gamma-ray burst continuum spectra have a well-known evolution as
the burst proceeds, both over the entire, often complex light
curves, and over individual pulses. The latter evolution is often
more spectacular. This evolution is generally characterized by an
overall softening of the spectra to lower energies with time. Two
specific correlations between observable quantities have been
found that describe this softening in a qualitative manner. The
first is the {\it hardness-intensity correlation} or HIC, which
relates the instantaneous hardness of the spectra and the
instantaneous energy flux $F_{E}$, within individual pulses. For
the decay phase of a pulse the most common behavior of the HIC is
\begin{equation}
F_{E}=\F0 (\Epk/\E00)^{\eta}, \label{HIC}
\end{equation}
\noindent where $\E00$ and $\F0$ are the initial values of the
peak of the $E F_{E}$ spectrum and the energy flux at the
beginning of the decay phase of each pulse, respectively. $\eta$
is the power law index\footnote{We follow the notation introduced
in \citet{RP}. Note that \citet{gol83} used $\gamma$ for the power
law index, which unfortunately is also a notation for the Lorentz
factor}. This is a general form for non-thermal energy flux
spectra that is largely independent of the emission mechanism
which gives $F$ an implicit time dependence because the peak
energy evolves with time. The original study by \citet{gol83}
found the power law index to vary between $1.5-1.7$ over the whole
GRB. Moreover, \citet{BR01} studied a sample of 82 GRB pulse
decays and found them to be consistent with a power law HIC in, at
least, 57\% of the cases and for these found $\eta = 2.0 \pm 0.7$.
Several physical emission processes are described by such a
relation, with the most prominent example being single particle
synchrotron emission with a constant magnetic field and a variable
electron Lorentz factor $\Gamma$. In this particular case, $E_{pk}
\propto \Gamma^{2}$ while $F \propto \Gamma^{2}$, so the HIC index
is $\eta=1$.

The second correlation is the {\it hardness-fluence correlation}
or HFC \citep{LK96} which describes the observation that the
instantaneous hardness of the spectra decays exponentially as a
function of the time-integrated flux, or fluence, of the burst:
\begin{equation}
    E_{pk} = E_{0} e^{-\Phi / \Phi_{0}}, \label{equ:LK96}
\end{equation}
where $\Phi(t)$ is the photon fluence integrated from the start of
the burst and $\Phi_{0}$ is the exponential decay constant.
\citet{LK96} found the HFC to be valid for 35 of 37 smooth GRB
pulses.
%uring their decay phase.
These findings were later confirmed by \citet{crid99} who studied
a sample of 41 pulses in 26 GRBs and found that the spectral
evolution in the majority of the pulses could be described by the
HFC. The physical interpretation of the HFC is not immediately
clear, but upon differentiation of equation (\ref{equ:LK96}), it
can be seen that it simply states that the rate of change in the
hardness is proportional to the luminosity of the radiating medium
(or, equivalently, to the energy density);
\begin{equation}
-dE_{pk}/dt  = -\frac{F_{N}E_{pk}}{\P0} \approx
-\frac{F_{E}}{\P0}.
\end{equation}

Here the last equivalency is only approximately true.  The energy
flux is actually defined as the integral of the photon energy
times the photon spectrum, $F_{E} = \int E N(E) dE$, where in this
case the integral would be carried over the BATSE energy band.  We
make the assumption that if $E_{pk}$ falls well within that band,
then the energy flux is approximately $E_{pk}$ times the
instantaneous photon flux, $F_{N}E_{pk} \approx F_{E}$. Since
$E_{pk}$ is defined to be the peak of the $E F_{E}$ spectrum, it
represents the energy of most of the arriving photons and
therefore this equivalency is robust as long as $E_{pk}$ does not
evolve outside of the BATSE window.  We also make the assumption
that the evolution of the low-energy power law index $\alpha$
should not effect our results except for pulses with an extreme
amount of evolution.  This assumption is supported by the findings
of \citet{crid99} and \citet{BR01} who avoided making the
assumption that $F_{N}E_{pk} \approx F_{E}$ by directly using
photon flux to test the HFC and HIC relations.  Using independent
methods, they show the approximation to be valid for a large
sample of GRBs.

Liang and Kargatis originally proposed that the HFC could be the
result of a confined plasma with a fixed number of particles
cooling via $\gamma$-radiation.  This type of exponential decay of
the break energy with photon fluence that is seen in the HFC is
expected if the average energy of the emitted photons is directly
proportional to the average emitting particle energy such as in
thermal bremsstrahlung or multiple Compton scattering, although
this interpretation is not unique.

\section{Analytic Light Curve Profile} \label{sec:decay}

We now follow \citet{RS00} to find an analytical description of
the energy-flux decay-profile. Note that in their original
description they used the photon flux, and we will here
reformulate their results in terms of the energy flux instead,
which makes a physical interpretation easier. Combining the
empirical relations, the HIC and HFC, gives the following
differential equation governing the spectral evolution

\begin{equation}
    \dot{E}=-\frac{F_0}{\phi_0 E_0^{\eta}} E^\eta
\end{equation}

\noindent The solution gives an expression for the energy flux
during the decay phase of a pulse

\begin{equation}
F(t) = \left\{ \begin{array}{ll}
            \F0 \left(1+\frac{(\eta-1)t}{T} \right)^{-\eta/(\eta-1)}
& \mbox{if $\eta
\neq 1$ }\\
            \F0 e^{-t/T} & \mbox{if $\eta = 1 $ ,} \end{array} \right.
\label{Flux}
\end{equation}

\noindent and correspondingly the peak of the {\bf $E F_{E}$}
spectra follow

\begin{equation}
\Epk(t) = \left\{ \begin{array}{ll}
            \E00 \left(1+\frac{(\eta-1)t}{T} \right) ^{-1/(\eta-1)}&
\mbox{if $\eta \neq
1$ }\\
            \E00 e^{-t/T} & \mbox{if $\eta = 1 $ ,} \end{array} \right.
\label{Et}
\end{equation}

\noindent where $T \equiv \P0 \E00/\F0$. Note that this gives the
possibility to measure $\P0$ directly from the light curve with
the knowledge of $\E00$. Introducing $d \equiv \eta/(\eta-1) $
($d$ as in the asymptotic {\it d}ecay of the energy flux) we can
describe the $E_{pk}(t)$ and the $F(t)$ decays as

\begin{equation}
F(t) =   \F0 \left(1+\frac{t}{T(d-1)} \right)^{-d}\label{e7}
\end{equation}

\begin{equation}
\Epk(t)=   \E00 \left(1+\frac{t}{T(d-1)} \right) ^{1-d}\label{e8}
\end{equation}

\noindent for the $\eta \neq 1$ case.  Note that this
representation is completely model-independent and is based solely
on the empirical HIC and HFC relations that describe GRB's
spectral evolution.  As will be shown in the following sections,
specific models, such as relativistic kinematics, make specific
predictions regarding the value of the HIC power-law index and
hence the value of decay constant $d$.

\subsection{Inclusion of the Rise Phase} \label{sec:rise}

The analytic profile derived above describes only the decay phase
of a pulse, since that is the portion of the GRB time history for
which the HIC is often a power law (eq. [3]) and the HFC follows
equation (4). This introduces an ambiguity when fitting equations
(\ref{e7}) and (\ref{e8}) to the data, since the power law index
$d$ and the time constant $T$ are coupled and are therefore not
well constrained by fitting. Furthermore, choosing the moment when
the pulse changes from the rise to the decay phase is somewhat
subjective.  In order to eliminate these ambiguities, a rise phase
must be supplemented to equations (9) and (10) in order to produce
a complete description of the FRED profile.  This also allows for
additional parameters such as the pulse rise time and asymmetry
values to be measured.

We first note the fact that in most physical models both the peak
of the energy spectrum and the luminosity are proportional to the
random Lorentz factor of the shocked electrons to some power
\citep{RB79}:

\begin{equation}
\Epk (t) \propto \Gamma_{\rm r}^a (t) \cdot g(t),
\end{equation}

and

\begin{equation}
F(t) \propto \Gamma_{\rm r}^b (t) \cdot h(t).
\end{equation}

\noindent The functions $g(t)$ and $h(t)$ parameterize the unknown
time dependencies on particle densities, optical depth, magnetic
field and kinematics, etc. The correlation between hardness and
the energy flux can thus be described as

\begin{equation}
F(t) \propto \Epk ^{\eta'} \cdot f(t)
\end{equation}

\noindent with $\eta' = b/a$ and $f(t)=h(t) g(t)^{-\eta'}$. During
the decay phase of a pulse the power law relation dominates the
HIC, which has been elucidated in previous studies. However, the
HIC during the rise phase will be dominated by the unknown
function $f(t)$.

We will now assume two simple representations of $f(t)$ and
recapitulate the analytical discussion in the previous section to
find a description for the shape of the entire pulse.

\subsubsection{Power-Law Rise}

The simplest assumption one can make is that the rise phase is
dominated by a power-law increase of flux. This leads to a simple
prescription of the function

\begin{equation}
f(t) \propto t^r
\end{equation}
\noindent
 where the power-law index, $r$, stands for the {\it
r}ise. In the case of optically-thin synchrotron emission the
energy flux is given by $F_{E} \propto {E_{\rm pk}} {\tau_{T}} B$,
so the HIC exponent in equation (\ref{HIC}) is unity, $\tau_T$ is
the Thompson depth, and $B$ is the magnetic field. In this case
the explicit time dependence arises from the evolution of the
Thompson depth and the magnetic field strength. Our initial
assumption gives rise to a general form that should also emerge
when considering other non-thermal emission mechanisms.

To fully model the FRED light curve we have to define the time at
which the light curve peaks as $t=t_{\rm max}$ or $t_{\rm m}$ for
brevity. This allows us to define several new parameters,
$\Epk(t=0)=E_0$ and $\Epk(t=t_{\rm m})=E_{\rm m}$ and
$F(t=\tm)=\Fm$ and that $\taum \equiv \Fm /\Em \P0$.  The HIC can
then be written as:

\begin{equation}
F=\Fm  \left( \frac{\Epk}{\Em} \right)^{\eta} \left(\frac{t}{\tm}
\right)^r \label{ekv:HIC2}
\end{equation}

\noindent which combined with the HFC  now gives the differential
equation

\begin{equation}
\dot{\Epk}=-\frac{\Fm}{\P0 \Em^{\eta} \tm^r} \Epk^\eta t^r
\label{equ:HFC}
\end{equation}

\noindent Now, we define the exponent $d$ (for finite $r$) to
describe the asymptotic power law behavior of the light curve,
$F(t) \rightarrow t^{-d}$ as $t \rightarrow \infty$ which gives
$\eta = (d+r)/(d-1)$. We can now better define the peak of the
pulse to be at $t=\tm$ by letting ${\rm d} F / {\rm d} t (t=\tm)
=0$ and get the condition that for $\tm \neq 0$,$\taum= \tm
(d+r)/[(d-1)r]$. Therefore, for a simple power-law rise, the final
pulse shape and $E_{pk}$ evolution can be described by:

\begin{equation}
F(t)={\Fm}\left(\frac{t}{\tm} \right)^r \left( \frac{d}{d+r} +
\frac{r}{d+r}\left(\frac{t}{\tm} \right)^{r+1}
\right)^{-\frac{r+d}{r+1}} \label{F1}
\end{equation}

\begin{equation}
\Epk(t)= {\Em}\left( \frac{d}{d+r} +
\frac{r}{d+r}\left(\frac{t}{\tm} \right)^{r+1}
\right)^{-\frac{d-1}{r+1}} \label{equ:Rsol2}
\end{equation}

\noindent This four parameter model fully describes a single
pulsed GRB event with direct measurements of an observed light
curve yielding the values for $\Fm, \tm,r,d$.    This then gives
$\taum$, which in its turn, gives the product $\Em \cdot \P0$. As
$\Em$ can be measured from the $\Epk$ decay, $\P0$ can be deduced.
Example profiles with varying decay parameters for the power-law
rise-model are shown in Figure 3.

It must be noted that in this scenario (rise+decay), the HIC power
law index, $\eta'$, will correspond to both the implicit power law
behavior, $E^{\eta}$ as well as the explicit behavior, $t^r$. Here
$\eta'$ denotes the power law in the second description. The two
descriptions coincide as $r$ tends to $0$ and as $\E00 \rightarrow
\Em$ which gives
\begin{equation}
F=\frac{\Fm}{\left(  1 + \frac{t}{(d-1) \taum} \right)^{d}}
\end{equation}
and thus $T=\taum$ by identification. Note that $\tm$ tends to $0$
as $r$ approaches $0$. It is also readily seen that the asymptotic
behaviors as $t$ approaches $\infty$ are the same.

\subsubsection{Exponential Rise}

It is reasonable that the rise phase is connected to some
transient process, e.g., an initial decrease in optical depth,
increase in the number of energized particles, or the merging (or
crossing) of two shells. After this initial phase, the original
decay behavior as described by equations (\ref{e7}) and (\ref{e8})
% \S \ref{sec:decay}
should emerge.
%Such a physical scenario could be the following.
We therefore try the following prescription

\begin{equation}
f(t) \propto 1-e^{-t/\tr}
\end{equation}

\noindent where the time constant $\tr$ now represents the {\it
r}ise phase. This corresponds to $h(t) \propto g(t) ^{\eta}$ as $t
>> \tr$. We therefore have

\begin{equation}
F=\Fm  \left( \frac{\Epk}{\Em} \right)^{\eta}
\frac{1-e^{-t/\tr}}{1-e^{-\tm/\tr}} \label{equ:F2}
\end{equation}

\noindent which gives the differential equation

\begin{equation}
\dot{\Epk}=-\frac{\Fm \Epk^\eta (1-e^{-t/\tr})}{\P0 \Em^{\eta}
(1-exp(-\tm/\tr))} \label{equ:diff2}
\end{equation}

{\noindent}In the same manner as before we define the decay index
$d$ by requiring $F(t) \rightarrow t^{-d}$ as $t \rightarrow
\infty$. This gives $\eta=d/(d-1)$. Solving the differential
equation (\ref{equ:diff2}) and combining with equation
(\ref{equ:F2}) we find

\begin{equation}
F(t)=\frac{A_0 (1-e^{-t/\tr})}{(1+(t+\tr e^{-t/\tr})/\td)^d}
\label{F2}
\end{equation}

\begin{equation}
\Epk(t)=\frac{A_1}{(1+(t+\tr e^{-t/\tr})/\td)^{(d-1)}}
\end{equation}
\noindent
 where $A_0$ and $A_1$ are analytical functions of $d,
\td, \tr, \P0, \Fm, \Em$ and $\tm$. There is no analytical
expression for the peak time $\tm$, but it can be solved for
numerically if the other parameters are known. Equation (\ref{F2})
describes the pulse by four parameters: $A_0, \tr, \td,$ and $d$.
This description coincides with the decay function (eqs.
[\ref{e7}] and [\ref{e8}])
%(\S \ref{sec:decay})
when $\tr \rightarrow 0$ or $t \rightarrow  \infty$ with $\td
\rightarrow (d-1) T$ (and $\Em, \Fm \rightarrow \E00, \F0$). Note
that in this description there are two explicit time constants
$\tr$ and $\td$, while in equation (\ref{F1}) the decay constant
is an expression of $r$ and $\tm$. For a sharp rise phase the
explicit time dependence will disappear. This is, however,  not
the case for equation (\ref{F1}).

\section{Data Analysis} \label{sec:analysis}

For the purposes of this study, we utilize 64ms count data
provided by the BATSE instrument onboard the {\it CGRO}
spacecraft. These data were gathered by BATSE's Large Area
Detectors (LADs) which provide discriminator rates with 64 ms
resolution from 2.048 s before the burst to several minutes after
the trigger \citep{fish94}. The discriminator rates are gathered
in four broad energy channels covering approximately 25-50,
50-100, 100-300, and 300 to about 1800 keV allowing for excellent
count statistics since the photons are collected over a wide
energy band.  For our analysis, we combine the data from all four
channels in order to study the "bolometric" light curve profile.
We chose to study the count flux light curves rather then directly
using the energy flux because of the spectral fitting that is
trespassory to obtain energy flux measurements. Although BATSE's
spectroscopic detectors (SDs) can gather enough data to produce
spectra on a 0.128 s timescale, the data usually has to be
integrated to get an acceptable fit to a given spectra model. This
reduces the resolution of the energy flux time history to a level
that is unacceptable for a four or five parameter fit.
Furthermore, \citet{br} have shown that, empirically, there is a
power-law relation between the count flux and the energy flux,
with an index clustering around $1$, which supports the use of the
count light-curves for our study of the pulse shapes. The
background is, in general, fitted to a quadratic polynomial using
1.024 s resolution background data that are available from 10
minutes before the trigger to several minutes after the burst. The
data along with the background fit coefficients were obtained from
the CGRO Science Support Center (GROSSC) at Goddard Space Flight
Center through its public archives.

We selected bursts from the entire BATSE catalog with the criteria
that the peak flux be greater than 1.0 photons cm$^{-2}$ s$^{-1}$
on a 256 ms timescale. Of these, we selected bursts that exhibited
clean, single-peaked events, or in the case of multi-peaked
bursts, pulses that were well distinguished and separable from
each other.  We limited the bursts to events with durations longer
than 2 s (full width half max - FWHM), primarily because this
study focuses on the properties of long GRB events. There are
strong indications that short bursts with $T_{90} < 2.6$ s have
different temporal behaviors compared to long bursts and that they
may actually constitute a different class of GRBs \citep{norRome}.
Therefore, in this study we are primarily interested in the shape
and temporal properties of long GRB events. In the search we did
not assume any functional description of the pulse profile, i.e.
pulses that did not look like FREDs were also selected, in order
to eliminate any preconceived idea of what the fundamental pulse
shape should be. We note that we do not know what the clean,
"generic" pulse actually looks like, and some noise, or
substructure, in the comoving light curve can be expected, for
instance, if the surface brightness of the shell is
non-homogeneous and/or if the density profile of the shell,
through which the shock is traversing, has a complicated radial
profile. Therefore, we adopt the following criteria. If a
significant substructure on top of a smoothly fitted pulse is
within a time interval of 1/4 the FWHM of the main pulse peak and
it contains more than  15 $\%$ of the total flux in the main
pulse, the pulse structure is rejected.  If the substructure
occurs in a time interval of more than twice the FWHM of the
primary pulse, then the pulse is treated as an independent event
and is fitted separately. We believe that this limits the possible
contamination of the sample with pulses that are actually a
composite of overlapping emission episodes. The final sample
consisted of 76 pulses within 68 bursts.  This sample differs from
those examined in previous studies such as \citet{nor96} because
they primarily represent separable pulses that exhibit long decay
tails that can be followed to background. Therefore, no attempt
was made to perform deconvolution fits of overlapping pulses, for
the purpose of obtaining robust measurements of the asymptotic
decay of each pulse, which is not possible for overlapping regions
of activity. For more details on the sample selection, see F. Ryde
$\&$ D. Kocevski (in prep.). These bursts are presented in
Table~1, and are denoted by both their GRB names and BATSE trigger
numbers. The pulses within multi-peaked bursts are distinguished
by their $t_{max}$ value.

All of the background-subtracted light-curves were fitted with the
power-law rise- (eq. \ref{F1}) and the exponential rise- (eq.
\ref{F2}) functions via $\chi^{2}$ minimization. To aid in this,
we developed a graphical IDL fitting routine which allowed the
user to set the initial pulse attributes manually before allowing
the fitting routine to converge on the best fit model. The
convergence routine utilized the Marquardt technique to minimize
the difference between the model and the time profile by adjusting
the model parameters and tracking the resulting effects on the
$\chi^{2}$. In the case of the power-law rise-function, the free
parameters were: peak flux $f_{max}$, position of peak flux
$t_{max}$, the rise index $r$, and the decay index $d$. Most of
the GRBs had significant emission prior to the BATSE trigger time,
so we introduced a fifth parameter $t_{0}$ which measures the
offset between the start of the pulse and the trigger time. The
initial value of $t_{0}$ was typically set to where the light
curve would rise above 10$\%$ of the background, helping to
eliminate the ambiguity associated with the start of the pulse. In
the case of the exponential rise-function, the free parameters
were: pulse amplitude $A_{0}$, rise time constant $\tau_{r}$,
decay time constant $\tau_{d}$, the asymptotic decay power law
index d, and the trigger offset $t_{0}$.

\section{Results} \label{sec:results}

Presented in Figure 4 are four type-fits to the power-law
rise-function (eq. \ref{F1}).  These pulses represent a good
example of the clean separable FRED pulses chosen for this
analysis and the degree of acceptability of $\chi^{2}$ for our
fits.  Overall the power-law rise-function appears to give a
better fit to most of the pulses compared to the exponential
rise-function (eq. \ref{F2}). This can be seen in the spread of
the respective $\chi^2$ distribution for our entire sample shown
in Figure 5, with the power-law rise-model giving a mean
$\chi^{2}$ of 1.16 with a standard deviation of 0.19.  This is
most likely due to the fact that the power-law function allowed
for the modelling of both concave and convex rise profiles,
whereas the nature of the exponential function is limited to
modelling concave shapes. This is a major constraint for the
latter model since many GRB pulses that are arguably single
emission episodes are seen to have concave rise profiles, at least
in the initial stage.

The narrow distribution of $\chi^{2}$ values also indicates that a
power law is sufficient to model the decay portion of the pulse
light curve.  According to \citet{Fen96}, this would not be the
case if evolution of the low energy spectral index $\alpha$
greatly affected the resulting light curve.  As discussed in $\S
1$, they show that the decay phase should scale roughly as the
power of the low energy spectral index $T^{-\alpha-1}$. If
$\alpha$ exhibited a large range of evolution during the decay
phase, then a single power law would no longer be adequate to
model the temporal behavior of the light curve.  The $\chi^{2}$
distributions support the assumption that the evolution of the low
energy spectral index does not significantly alter our results.

\subsection{Rise and Decay Timescales} \label{sec:risedecay}

The primary pulse attributes that can be derived from fitting the
time profiles are related to the pulse rise and decay rates.  In
the case of the power-law rise-function, these consist of the
asymptotic rise and decay power-law indices $r$ and $d$ described
in $\S$2.3.  The resulting  $r$ and $d$ parameter distributions
for our entire sample are shown in Figure 6. The rise index
appears to be the tighter of the two, with a mean value of
$\langle r \rangle = 1.49 \pm 0.25$ and a standard deviation of
0.63. The distribution of the decay index  is somewhat broader
with a mean value of $ \langle d \rangle = 2.39 \pm 0.12$ with a
standard deviation of 0.76. The exponential-rise function has no
corresponding rise index, but rather the rise timescale is
associated with the time constant $\tau_{r}$ in equation
(\ref{F2}).  The fit results show that the rise timescale covers a
very broad range of values with a median value of $ \langle \tr
\rangle = 2.40 \pm 0.10$ and a standard deviation of 2.50. The
asymptotic decay index is similar to that found with the power law
rise model except that it has a much broader distribution with a
median value of $ \langle d \rangle =2.75 \pm 0.12$ with a
standard deviation of 1.50.

\subsection{Pulse Asymmetry} \label{sec:pasym}

A measure of the average pulse time-asymmetry can be easily
constructed using the fit parameters obtained from both models.
Several authors have studied burst asymmetry properties in the
past \citep{nor96,Fen96} -  and found interesting correlations
between the pulse asymmetries and other temporal properties.
\citet{nor96} measured asymmetry values for some 400 fitted pulses
and reported a general trend that narrower pulses tend to be more
symmetric and have harder spectra.  This is often referred to as
the GRB pulse paradigm. Here we define the asymmetry as the
fraction of the FWHM belonging to the rise phase of the pulse to
the fraction belonging to the decay phase (see Figure 7 for an
illustration). This description is independent of the model
parameters used to fit the pulses and hence allows the asymmetry
to be measured for both of the functions utilized in this analysis
in a comparative manner. The resulting asymmetry distribution for
the power-law rise-model is remarkably tight with a median value
of 0.47 $\pm$ 0.08 and a standard deviation of 0.09. The
exponential rise model provides similar results yielding a median
value of 0.45 $\pm$ 0.07 and a standard deviation of 0.07.

Alternatively, the asymmetry can be defined using the $T_{90}$ of
the pulse instead of the FWHM.  The rise phase would then be
defined as the interval from when where 5$\%$ of the pulse counts
are accumulated till the time of peak flux as measured by the
model parameter $t_{max}$. Similarly, the decay phase is simply
the time between the peak flux till when 95$\%$ of the pulse
counts are collected. Note that here we are using $T_{90}$ for the
individual $pulses$, not the entire burst which is the common
practice. Although this description is straightforward, it is
somewhat more problematic when dealing with multi-pulsed bursts
because subsequent pulses will routinely begin before the signal
from the first pulse falls to within 5$\%$ of the background.  It
is presumed that the decay phase of the pulse is hidden under the
rise phase of the following pulse, therefore in these cases the
decay time is only a lower limit to the true length of the decay
phase of the pulse.  This occurred in 10$\%$ of the pulses
observed, so the effect on the overall distribution should be
minimal.  The resulting median asymmetry values using the $T_{90}$
definition for both the power law and exponential models are 0.30
$\pm$ 0.04 and 0.21 $\pm$ 0.03 respectively. The resulting
asymmetry distributions for the power-law rise-function for the
FWHM definition is plotted in Figure 8.

Surprisingly, we find no evidence of a connection between the
pulse asymmetry and the pulse duration ($T_{90}$) for the
individual FRED pulses. Likewise, we find no significant
correlations between the pulse asymmetry and the pulse width
(FWHM).  The lack of any significant correlation can be seen in
Figure 9 and measured quantitatively by the use of the Spearman
rank-order statistic which provides a robust and convenient means
of evaluating the statistical significance of a given correlation.
A Spearman rank order analysis on the data shown in Figure 9
yields a correlation coefficient of $R = 0.057$, where $R = 1$
represents a 1:1 correspondence. Both of these conclusions are
contrary to the pulse paradigm referred to earlier and the
disagreement is most likely due to the difference in samples used
in the two studies. The \citet{nor96} study examined complex
overlapping bursts for a majority of their data sample, while the
majority of our sample consist of single-peaked FRED bursts.  This
leads to the conclusion that most of the simple FRED pulses that
were examined in our sample are extremely self-similar,
independent of their duration and amplitude.

\subsection{Correlations}  \label{sec:corr}

We now examine several correlations between temporal properties
measured from the pulse profiles such as the FWHM, rise and decay
times and fluence values.  First, Figure 10 shows the pulse rise
time versus the pulse width, as measured by the FWHM, for our
entire sample.  The resulting linear correlation has a slope of
0.323 $\pm$ 0.01 with an associated Spearman rank order
correlation coefficient of $R = 0.981$ and is a direct consequence
of the tight asymmetry distribution discussed above. This same
effect is also observed between the time constant $\tau_{r}$
derived from the exponential rise model and the pulse width.  Here
again, the rise time constant, which is an alternative way of
measuring the characteristic rise timescale, is linearly
correlated to the FWHM. Note that the FWHM values plotted in
Figure 10 are not normalized to any timescale such as the pulse
duration, reasserting that the rise time is a constant fraction of
the overall pulse width independent of the duration of the pulse.

Another correlation born from the data is the relationship between
the normalized rise time and the decay power-law index, which is
shown in Figure 11 and has a correlation coefficient of $R = 0.799
$. Here the rise time is normalized to the FWHM of the pulse, so
the longer the rise time is, relative to the FWHM, the steeper the
resulting power-law decay of the pulse envelope becomes. This is
very interesting since the rise time is thought to be proportional
to the dynamic or shell-crossing time of the pulse and is
generally believed to be independent of the subsequent fall of the
FRED profile.  A similar effect can be seen in Figure 12 where the
pulse asymmetry data are plotted versus the asymptotic decay
power-law index.  The degree of asymmetry is seen to be
proportional ($R = 0.831$) to the decay index in such a way that
pulses with steep decay profiles (large $d$) have a larger
rise-to-decay ratio. The latter correlation is to be expected
since, as $d$ tends to 0, the decay time approaches $\infty$ and
hence the asymmetry tends to 0. A similar correlation is not seen
with the rise power-law index.

\section{Discussion} \label{sec:discussion}

The fit results presented above indicate that the analytic
functions derived in \S 3 are very successful in describing the
overall profile of individual GRB pulses.  These descriptions
differ from those used in previous modelling surveys as they are
motivated by physical first principles within the general fireball
model with the requirement that the previously observed empirical
relations for the spectral evolution must hold. These descriptions
also build on the original \citet{RS02} analysis by including a
description of the rise phase of the pulse.  Of the two assumed
rise profiles presented in \S 2, a simple, power-law rise seems to
better describe the majority of the FRED pulses examined. This is
one of the simplest assumptions that can be made regarding the
dynamic (or shell crossing) phase of a GRB pulse and may point to
the explicit time-dependance of the parameters internal to the
shock such as Thompson depth or the magnetic field.  Hence, the
measured pulse parameters have the potential to be used to
diagnose pulse characteristics, such as bulk Lorentz factors,
$\Gamma$, shell radii, and thickness.

The decay power-law index appears to cover a wide range of values
with a median centered around 2.39 $\pm$ 0.12, which is 3.25
$\sigma$ steeper than the analytically derived value of $d=2$ for
pulse profiles created principally by relativistic effects due to
spherical symmetric shells.  Overall, 30 of the 77 pulses that we
analyzed had decay indices that were consistent (within 1$\sigma$)
with the predicted $d=2$ value, constituting 39$\%$ of our sample.
We note that the distribution of HIC indices, which were shown in
$\S 2$ to be directly related to the decay index, found by
\citet{BR01} also has a broad distribution, with several cases
differing substantially from $\eta=2$. There are several
mechanisms that can cause the decay envelope to deviate from the
predicted spherical scenario, but the most obvious would be the
breaking of local spherical symmetry. Both prolate and oblate
shell geometries as seen by the observer would create light curves
with resulting power-law indices that differ from the spherical
case.  In $\S 1$ we derived the shape that the light curve is
expected to take if the pulse profile were created purely by a
spherical shell.  In the case of an elliptical geometry, the time
delay between the arrival of off axis photons will be given by
$\Delta t = (R_0-R\mu)/c$, where $R_{0}$ is the radius of the
shell at $\theta=0$.  The radius $R$ of the shell is no longer
independent of the angle $\theta$, and is given by
\begin{equation} \label{rellipse}
R=a\sqrt{\frac{1-e^{2}}{1-e^{2}\mu^{2}}}
\end{equation}
where e is the eccentricity of the shell, a is the semi-major
axis, and $\mu=cos(\theta)$.  Substituting this back into $\Delta
t$, we find that
\begin{equation} \label{rellipse}
\Delta t=a-a\sqrt{\frac{1-e^{2}}{1-e^{2}\mu^{2}}\mu}
\end{equation}
Solving for $\mu$ and substituting into the Doppler boost
expression (equation \ref{doppler}) gives the Doppler profile
(Doppler boost as a function of time) for elliptical shells
\begin{equation} \label{doppler2}
{\cal D} (\Gamma,t)=
\frac{1}{\Gamma(1-\frac{\beta(-t+a)}{\sqrt{e^2t^2-2e^2ta-a^2}})}
%\frac{1}{\Gamma(1-\frac{\beta(-t+a)}{\sqrt{e^2t^2}}}}
\end{equation}
Following from the discussion in $\S 1$, the resulting evolution
of the spectral break-energy and light curve of the energy-flux
should follow $\Epk(t)=E'_{\rm pk}{\cal D}$ and $F_{bol}(t) =
F_{0} {\cal D}^{2}$, so the profile of the varying Doppler boost
should directly give rise to the time dependence of the $E_{pk}$
evolution and the pulse light curve.  The Doppler profile,
expressed in equation \ref{doppler2}, is shown for several
different prolate and oblate shock fronts in Figure 13, with the
corresponding shell geometry plotted as an inset. It shows that
the prolate shells (solid lines) have steeper Doppler profiles
resulting in asymptotic power law slopes which are greater than
the spherical case (thick solid line) whereas the opposite is true
for the oblate geometries (dashed lines). The extreme limit of the
oblate scenario is such that the shell becomes a parallel slab, in
which case for an unresolved source the Doppler profile approaches
zero and any pulse evolution is directly due to the shell's
intrinsic emission profile. Because of this effect, the observed
range of power-law decay indices may be explained by a simple
distribution of shell geometries, where the median decay index of
2.39 $\pm$ 0.12 would indicate that, in the context of
relativistic curvature, the majority of the analyzed pulses were
produced by shells with a degree of curvature greater than that
exhibited by a spherical shell.  Such conditions are not
unreasonable in the context of the fireball model and a simple
angular dependance of the bulk Lorentz factor can easily produce
an elliptical shock front that evolves as the shock propagates.
The evolution of the shock geometry, from oblate at early times to
prolate at later times, and a variation in the time at which the
shell "ignites" during this evolution, could account for the
distribution of decay indices.

\citet{Fen96} and \citet{FenRik} provide several other arguments
in support of the breaking of local spherical symmetry, the most
prominent being that the pulse photon light curves appear to
evolve uncoupled to the GRB spectra. They find that the GRB
spectra also evolve faster than the predicted rate of evolution
due to the simple boost factor that is expected from a spherically
symmetric shell. According to the investigation performed by
\citet{RP}, this result is consistent with our findings that the
average light curve profile also decays faster than predicted.  In
the context of their model, a range of relative sizes between the
comoving and the curvature timescales defined in $\S 1$ can result
in a variety of pulse shapes. Light curves that exhibit a
substantially different shape than that predicted by spherical
curvature are attributed to a scenario where the comoving
timescale dominates within a burst that already has an
intrinsically fast decay rate ($d>2$) of the comoving light curve.
This, according to their model, must also be manifested in the
spectral evolution in a corresponding manner.  Therefore, an
analysis of the spectral evolution rates and $\alpha$ values of
the pulses that vary the furthest from the predicted scenario of
$d=2$ may test how much these values affect the resulting light
curve profiles.  This research is currently being pursued (F.
Ryde,\& D. Kocevski in prep).

The asymmetry results are rather surprising and the reason for the
tight asymmetry distribution among such a variety of GRB pulses is
not immediately clear.  It would indicate that the dynamical and
angular timescales are not fully independent and that the observed
rate at which the shell becomes active is dependant on the decay
timescale and hence the curvature of the shell.  A situation can
be envisioned where a short comoving rise profile is delayed and
boosted to the same degree as the shell's intrinsic cooling
profile.  If both of these timescales are extremely short as
compared to the angular spreading timescale, then a relationship
can be produced between the observed rise and decay profile.  This
effect could produce a narrow asymmetry distribution and would
also predict a correlation between the rise timescale and decay
power-law indices, which is indeed observed in our sample.
\citet{RP} have discussed extensively the effect of varying
intrinsic emission profiles and angular spreading timescales on
the resulting light curve that is observed.  They find that the
resulting pulse asymmetry relies heavily on the convolution of the
intrinsic emission profile and the angular spreading timescale.
Therefore, future modelling of pulse timescales and shell
geometries, with the constraint that they produce the observed
asymmetry distribution, may yield clues to the intrinsic profile
of the emission episode.

\acknowledgments

We are grateful to the GROSSC at NASA/GSFC for providing the
HEASARC Online Service. We should also like to thank Dr. Vah\'e
Petrosian for his valuable discussions and insights.  D.K. wishes
to express his gratitude to the Department of Physics at Stanford
University for hospitality and also acknowledges the NASA GSRP
fellowship program for their support.  F.R. acknowledges financial
support from the Swedish Foundation for International Cooperation
in Research and Higher Education (STINT) and the Ludovisi
Boncomagni, n\'ee Bildt, foundation.

\clearpage

\begin{figure} \label{Fig:lightcurves}
\epsscale{1.0}
 \plotone{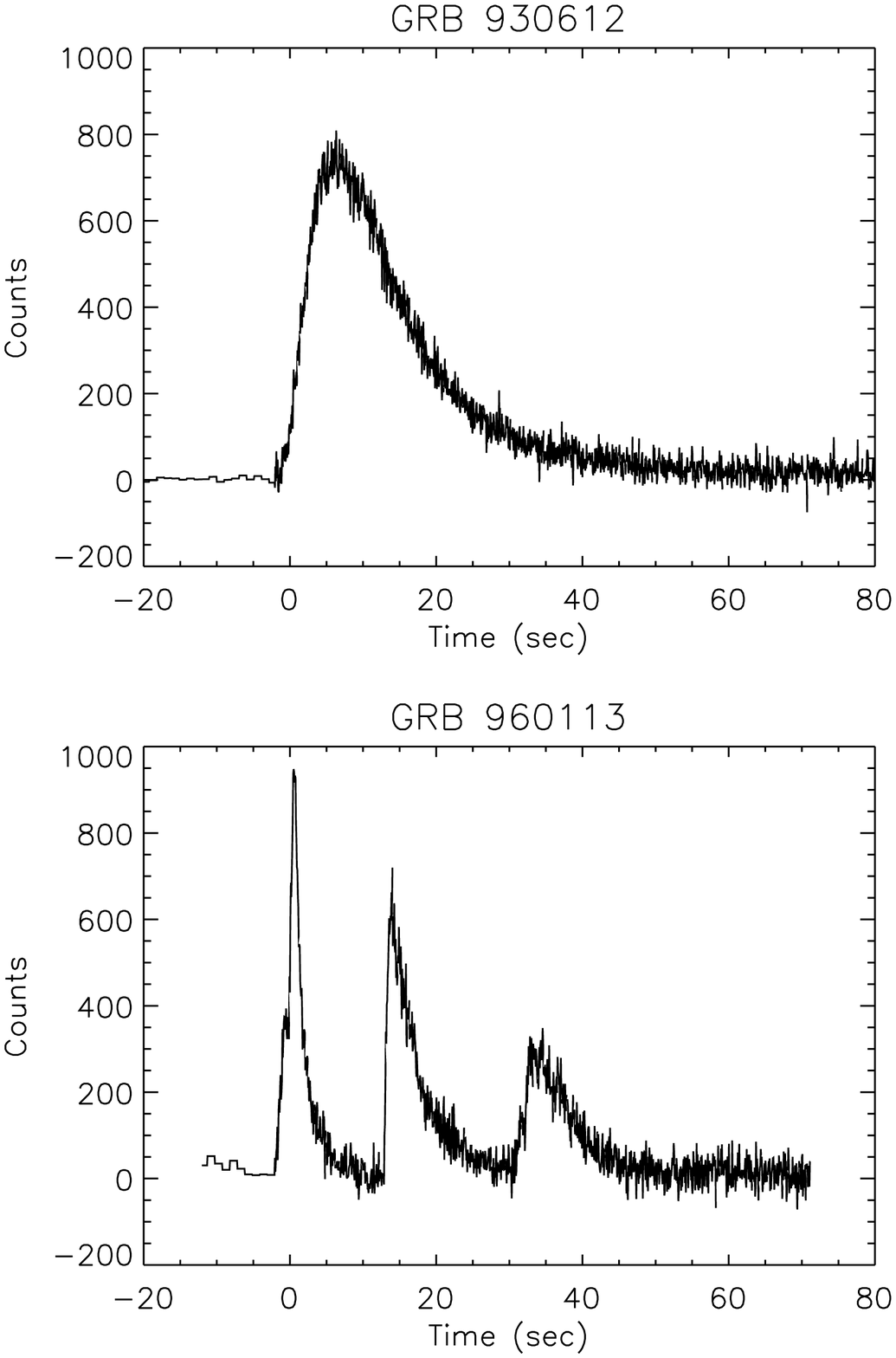}
 \caption{(a) Background
subtracted (64ms resolution) count light-curve (energy range: KeV)
for GRB 930612 (BATSE trigger 2387), a quintessential FRED pulse.
(b) Similar light-curve for GRB 960113 (trigger 4350) which
exhibits three separable FRED pulses.}
\end{figure}

\clearpage

\begin{figure} \label{Fig:curvature}
\epsscale{1.0}
 \plotone{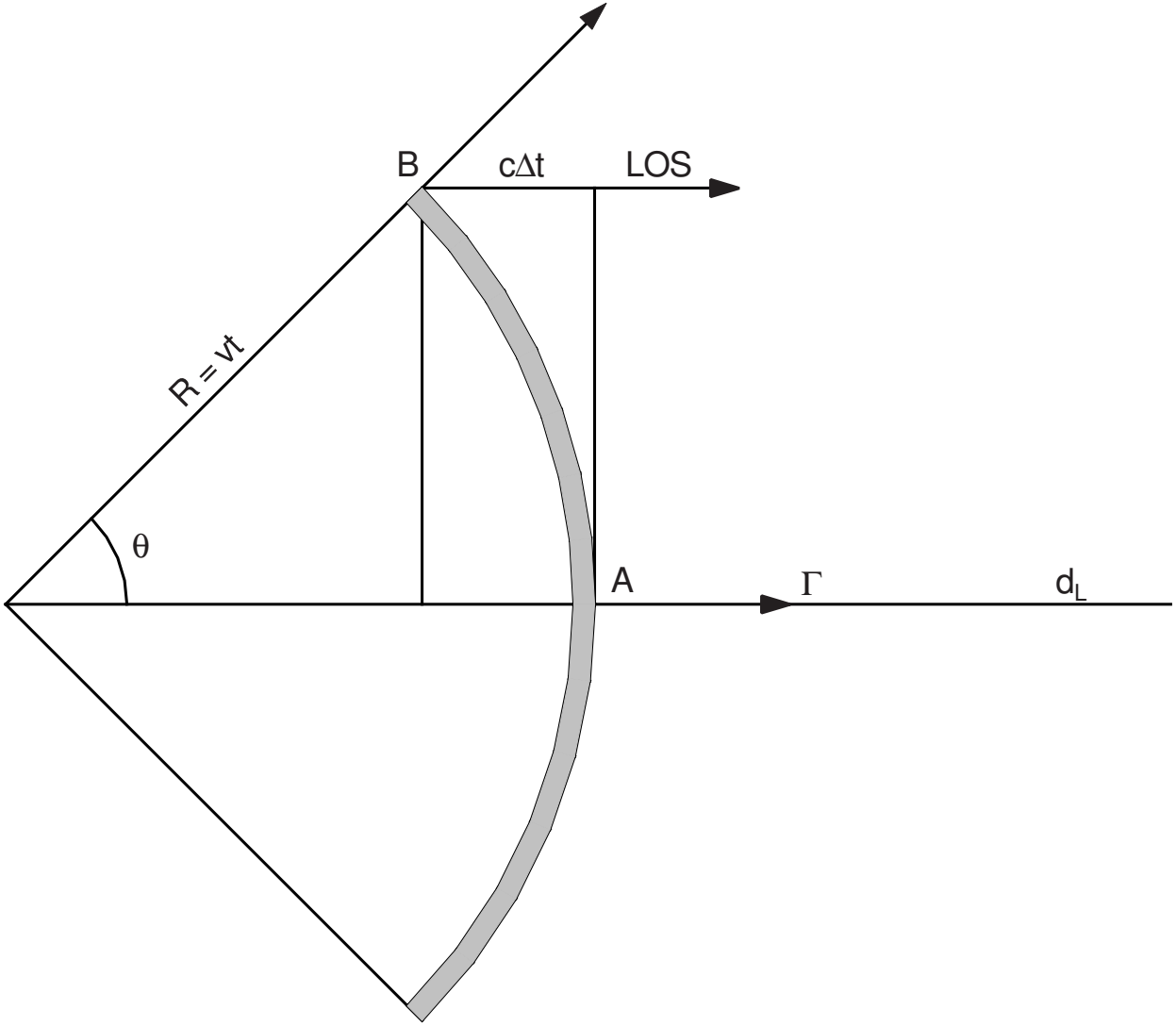}
\caption{(a) Schematic drawing of the visible part of the fireball
shell.  The photons from A are boosted by 2$\Gamma$, whereas the
photons from B are boosted by $\Gamma(1-\beta\cos{\theta})^{-1}$
and delayed by $\Delta t = R/c(1-cos{\theta}).$ Adapted from
\citet{RP}.}
\end{figure}

\clearpage

\begin{figure} \label{Fig:ricemodel}
\epsscale{1.0}
 \plotone{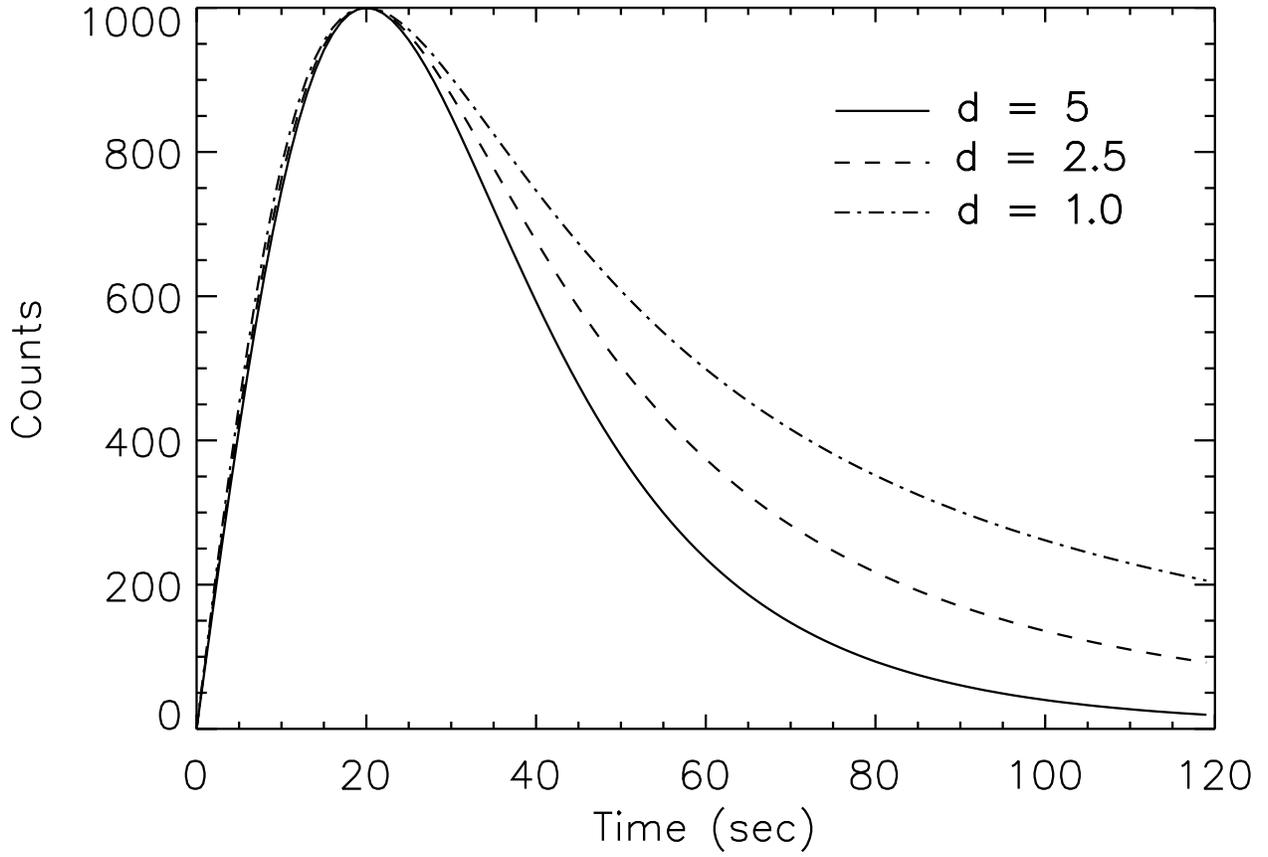}
\caption{Examples of profiles of the power-law rise-model for
three different power-law indices during the decay phase; $d=5$
(solid line), $d=2.5$ (dashed line), $d=1$ (dashed-dotted line).}
\end{figure}

\clearpage

\begin{figure} \label{fig:fits}
\epsscale{1.0}
 \plotone{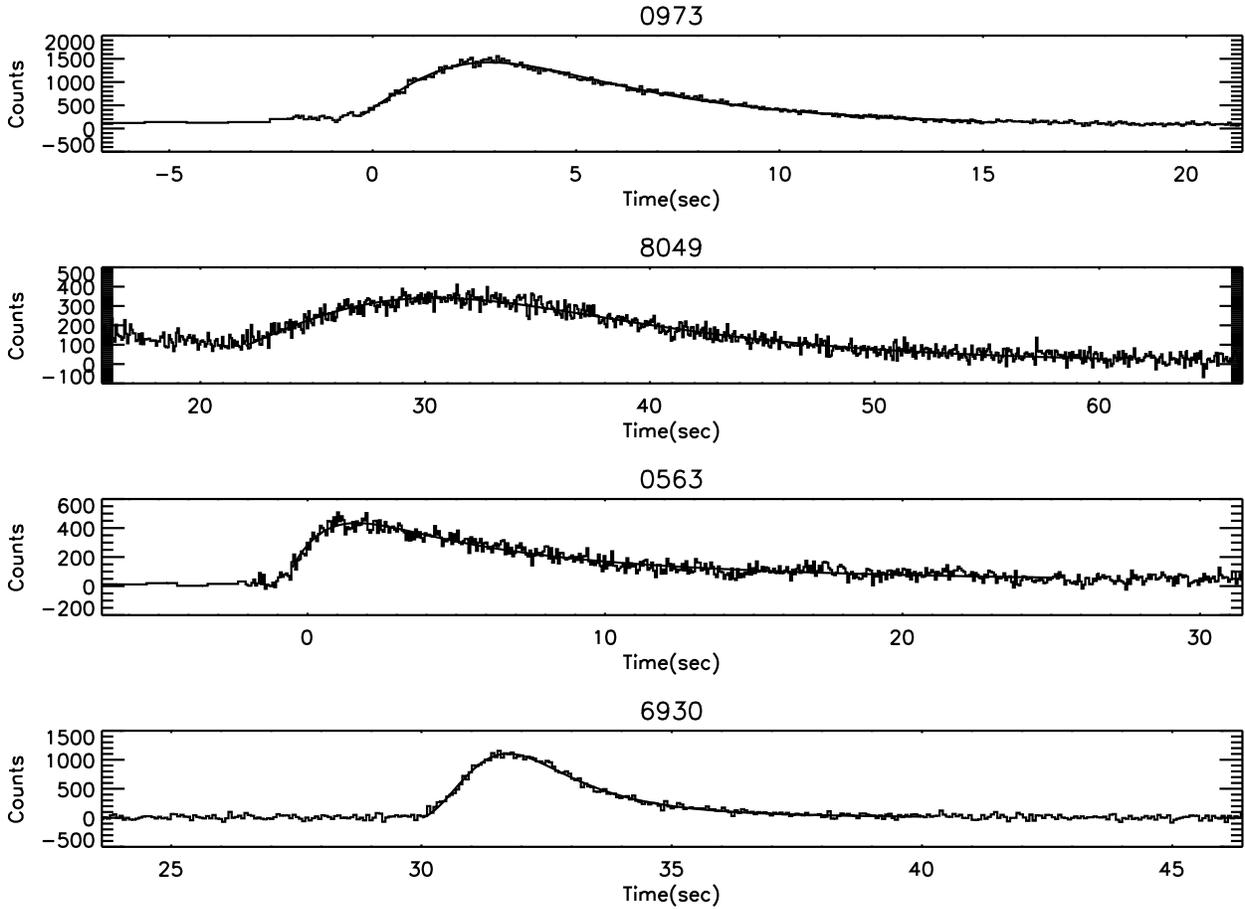}
\caption{Examples of fits to the power-law rise-model for GRB
911016b (trigger 907), GRB 000323 (8049), GRB 910721b (563), and
GRB 980718a (6930). See Table 2 for the associated fit
parameters.}
\end{figure}

\clearpage

\begin{figure} \label{fig:chisqr}
\epsscale{1.0}
 \plotone{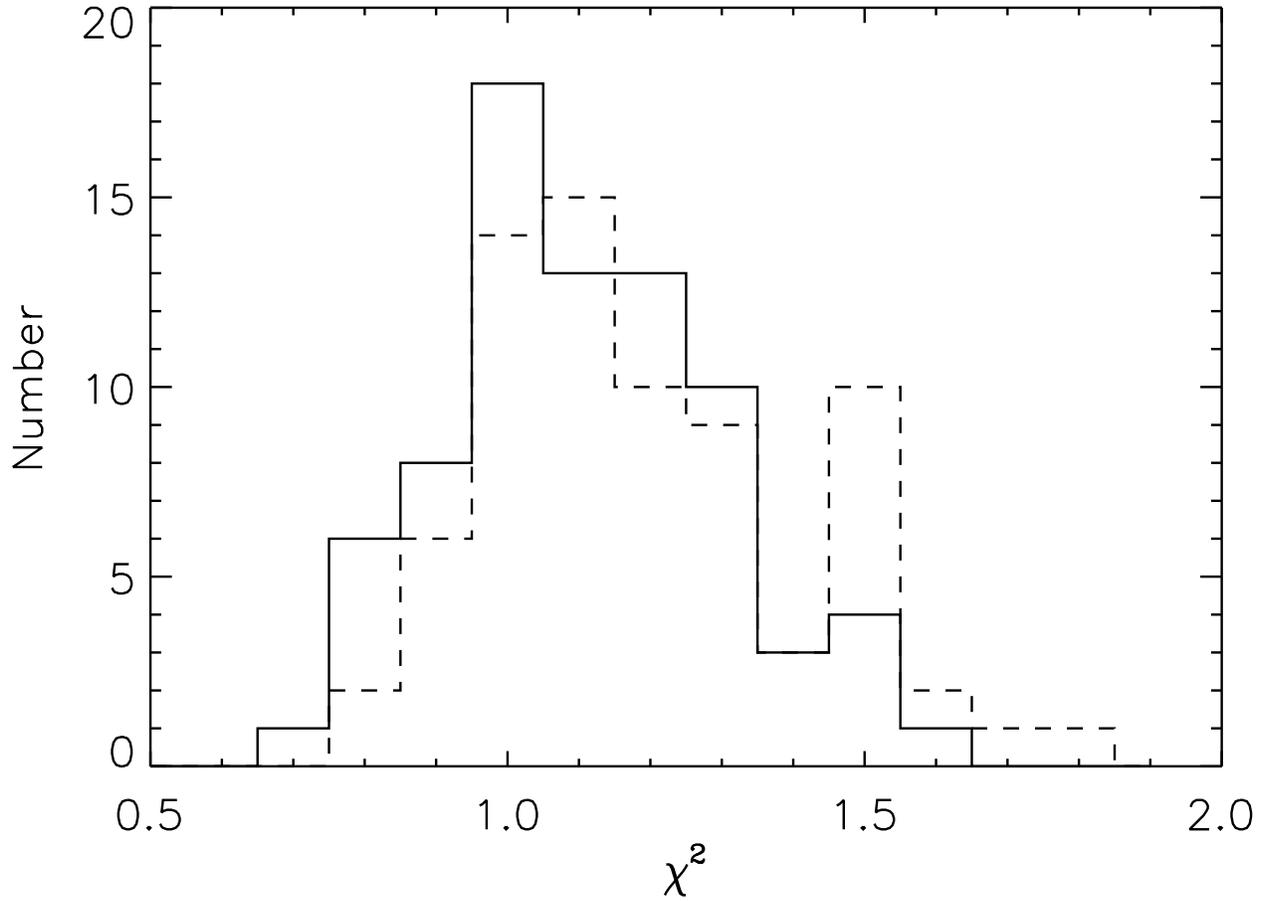}
\caption{$\chi^{2}$ distributions for the power-law rise-model
(solid line) and exponential rise-model (dashed line). A simple
power-law representing the rise-phase provides an overall better
fit to the data sample.}
\end{figure}

\clearpage

\begin{figure} \label{fig:rddist}
\epsscale{1.0}
 \plotone{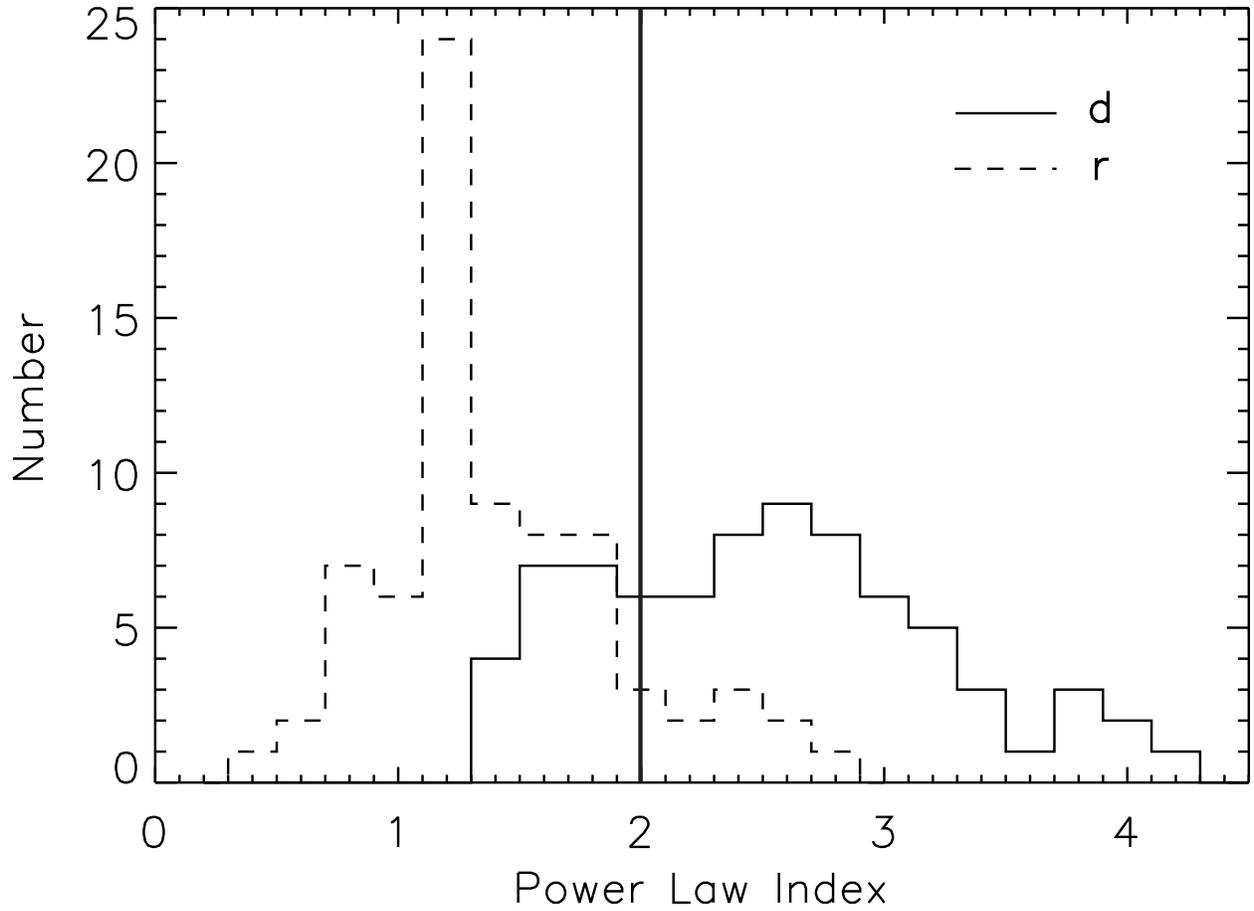}
\caption{Distribution of the rise (dashed line) and decay (solid
line) power-law indices for the power-law rise-model.  The thick
solid line represents the expected decay value if the pulse
profile were produced purely by curvature effects.}
\end{figure}

\clearpage

\begin{figure} \label{fig:asymplot}
\epsscale{1.0}
 \plotone{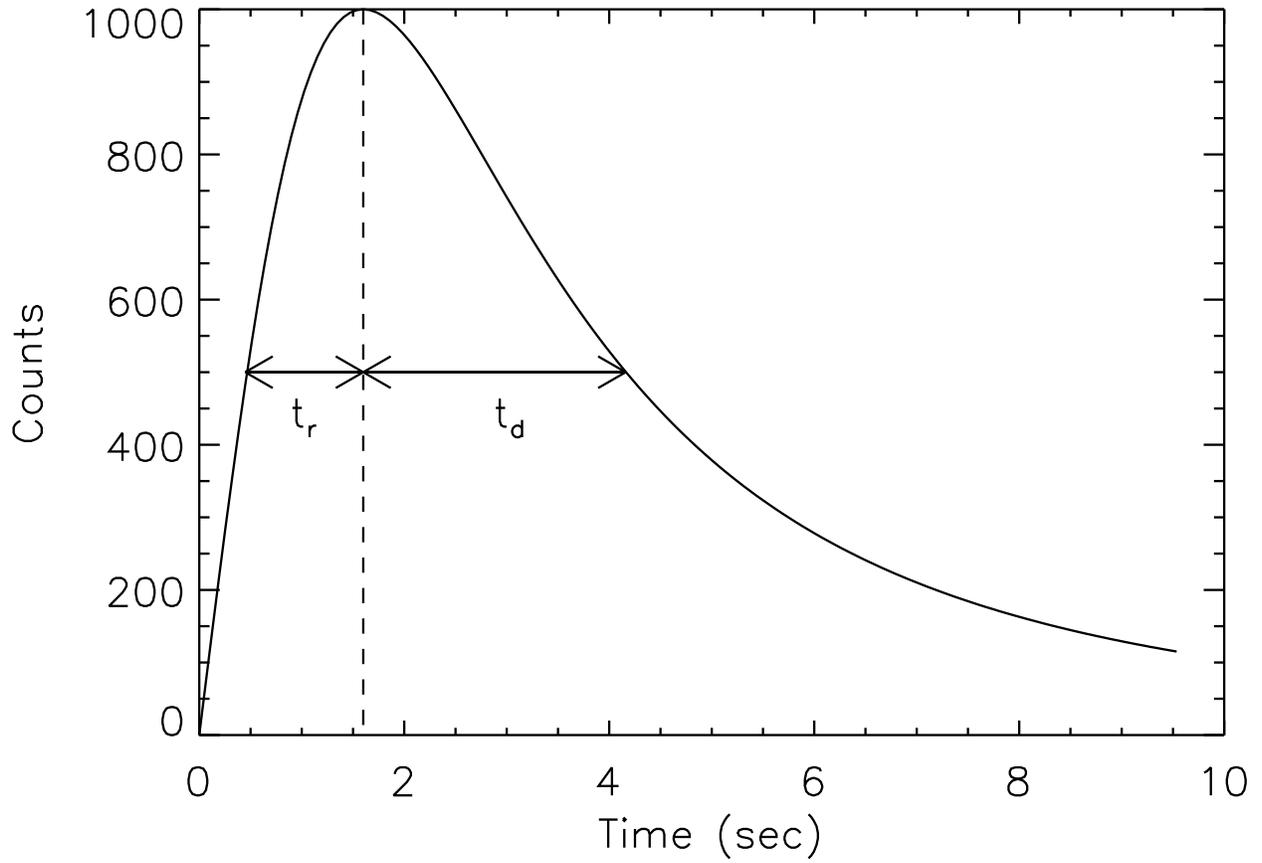}
\caption{Rise and decay times used to calculate the pulse
asymmetry are defined as the fraction of the FWHM before and after
the time of peak emission, ${t_{\rm max}}$.  This example has a
rise index of $r=1$ and a decay index of $d=2$.}
\end{figure}

\clearpage

\begin{figure} \label{fig:asymdist}
\epsscale{1.0}
 \plotone{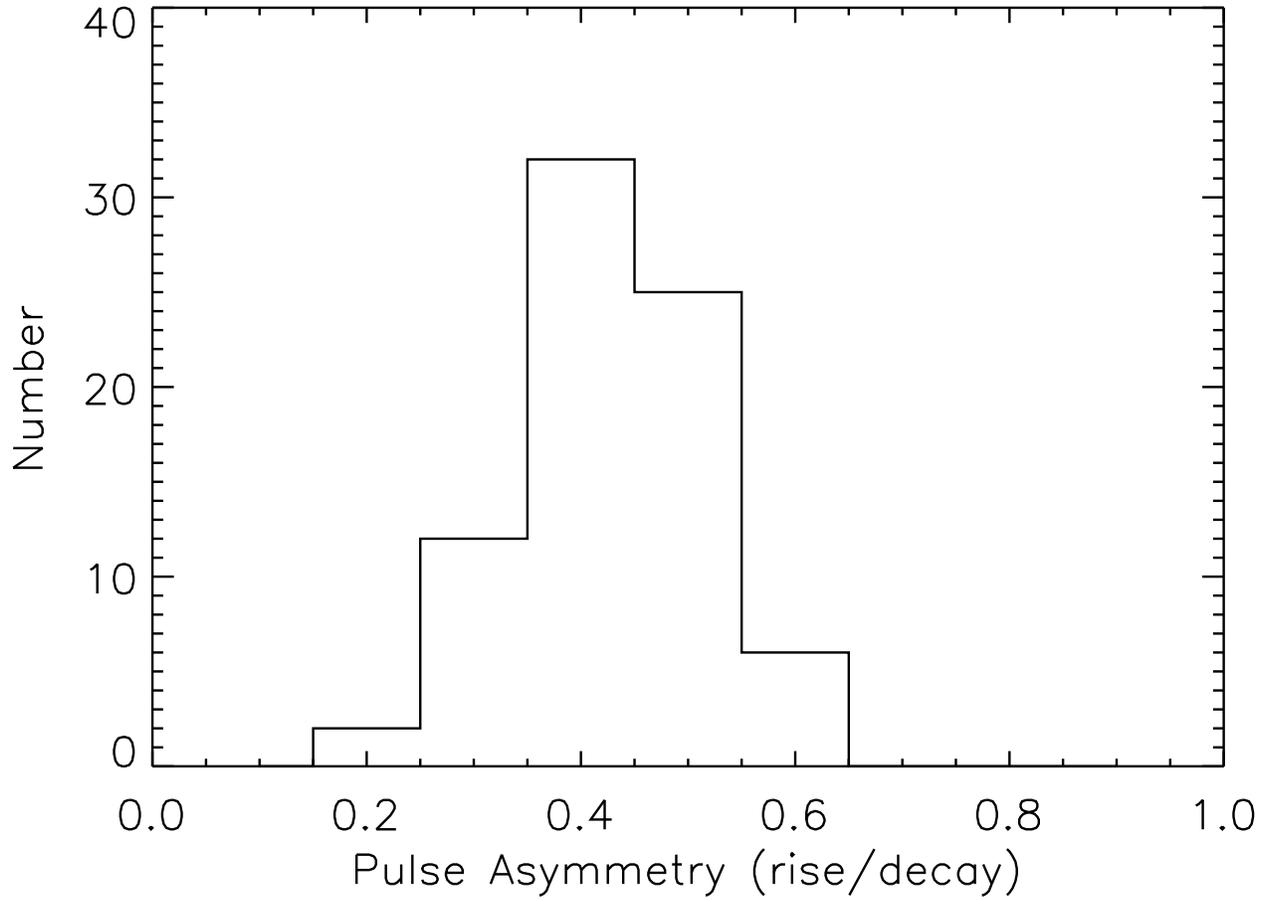}
\caption{Pulse asymmetry distribution for the power law rise model
as measured by taking the ratio of the rise to decay times, see
section \S 4.2 for a complete definition.  The distribution peaks
at 0.47 $\pm$ 0.08.}
\end{figure}

\clearpage

\begin{figure} \label{fig:t90}
\epsscale{1.0}
 \plotone{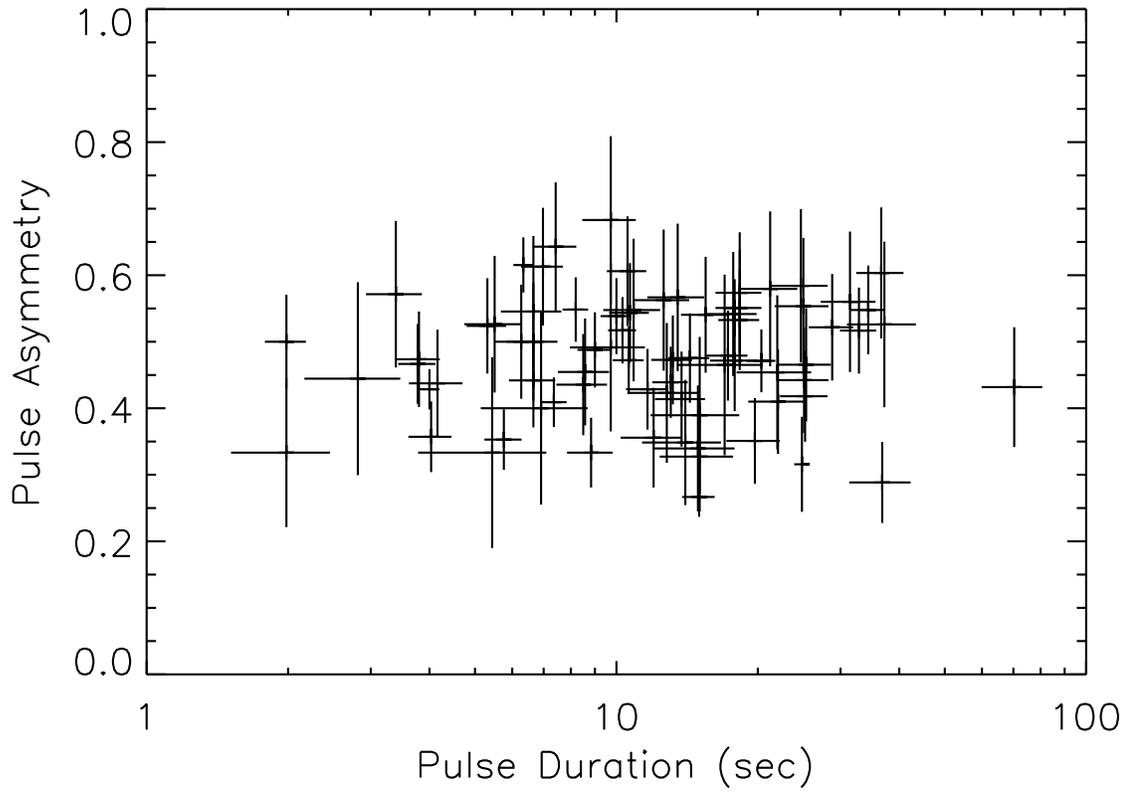}
\caption{Pulse asymmetry vs. the pulse duration ($T_{90}$) showing
no correlation between the length of the pulse to the ratio of its
rise and decay timescales.  The correlation coefficient is $R =
0.057$.}
\end{figure}

\clearpage

\begin{figure} \label{fig:risevsfwhm}
\epsscale{1.0}
 \plotone{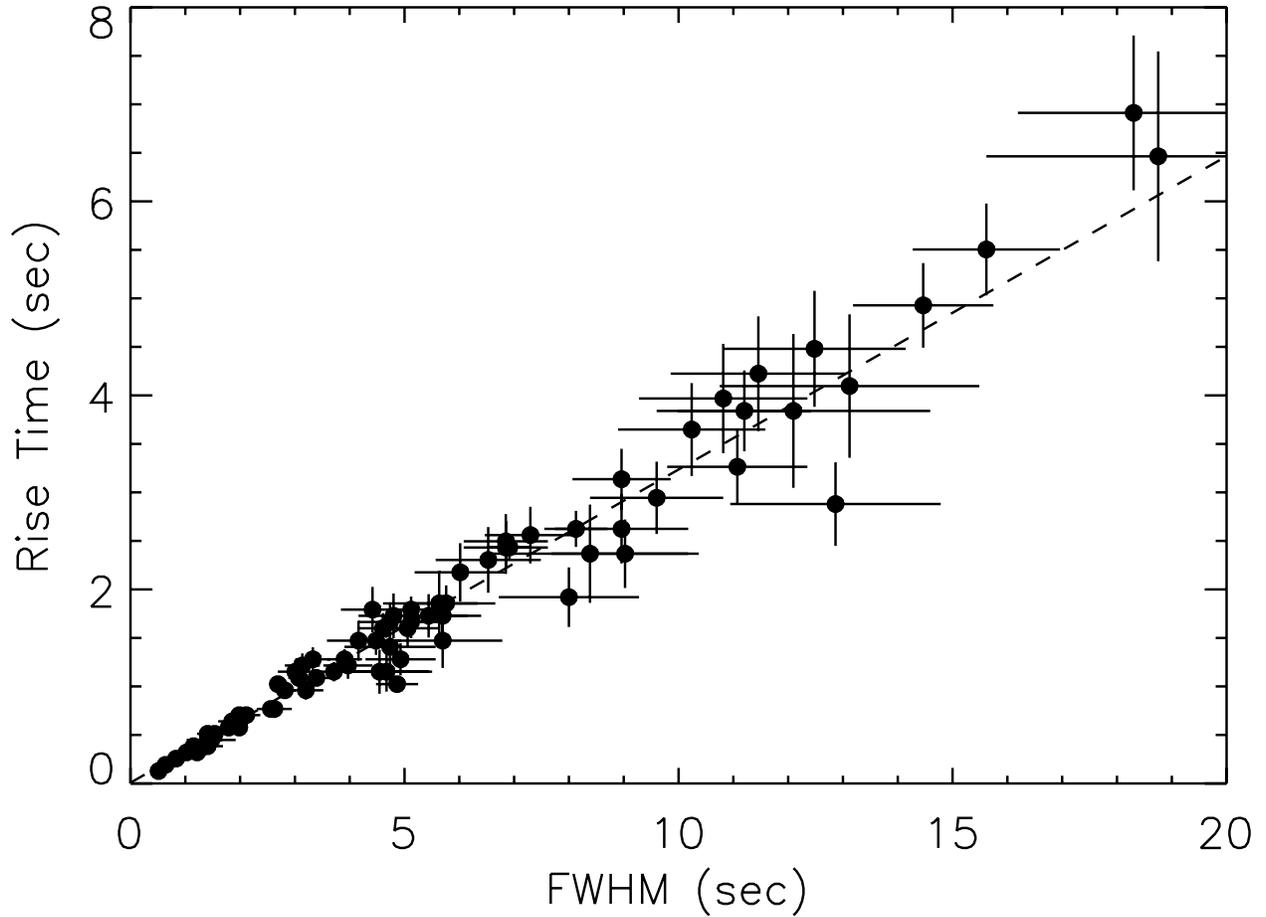}
\caption{Rise time vs. the FWHM as measured by the model with a
power-law rise.  The linear correlation is a manifestation of the
tight asymmetry distribution.  The correlation coefficient yields
a value of $R = 0.981$.}
\end{figure}

\clearpage

\begin{figure} \label{fig:dvsrise}
\epsscale{1.0}
 \plotone{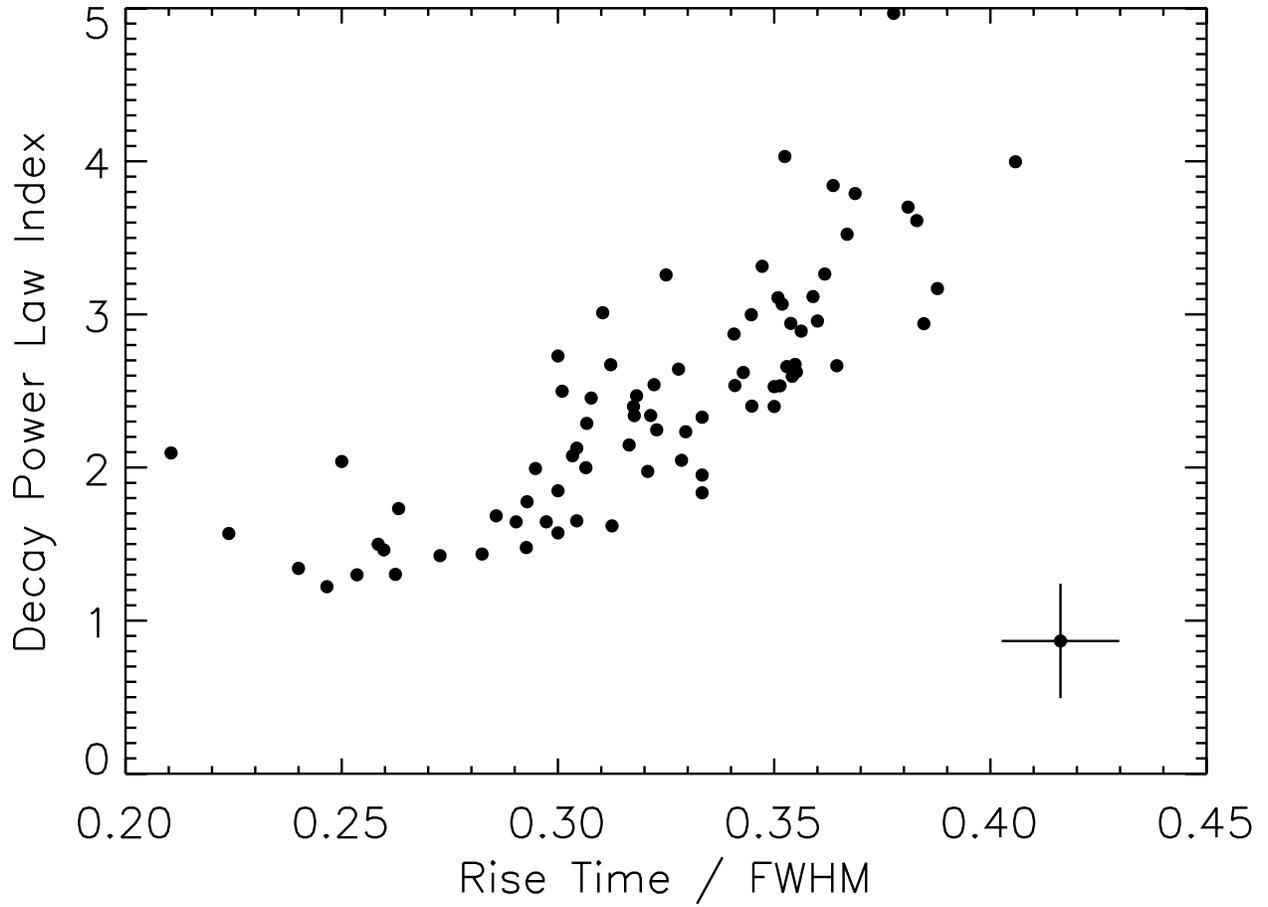}
\caption{Decay power-law index plotted vs. the normalized rise
time.  The time to peak is correlated to the hardness of the decay
profile with a correlation coefficient of $R = 0.799$. This again
is an effect of the narrow asymmetry distribution exhibited by the
pulses.  The average error is illustrated in the bottom right
corner.}
\end{figure}

\clearpage

\begin{figure} \label{fig:dvsasym}
\epsscale{1.0}
 \plotone{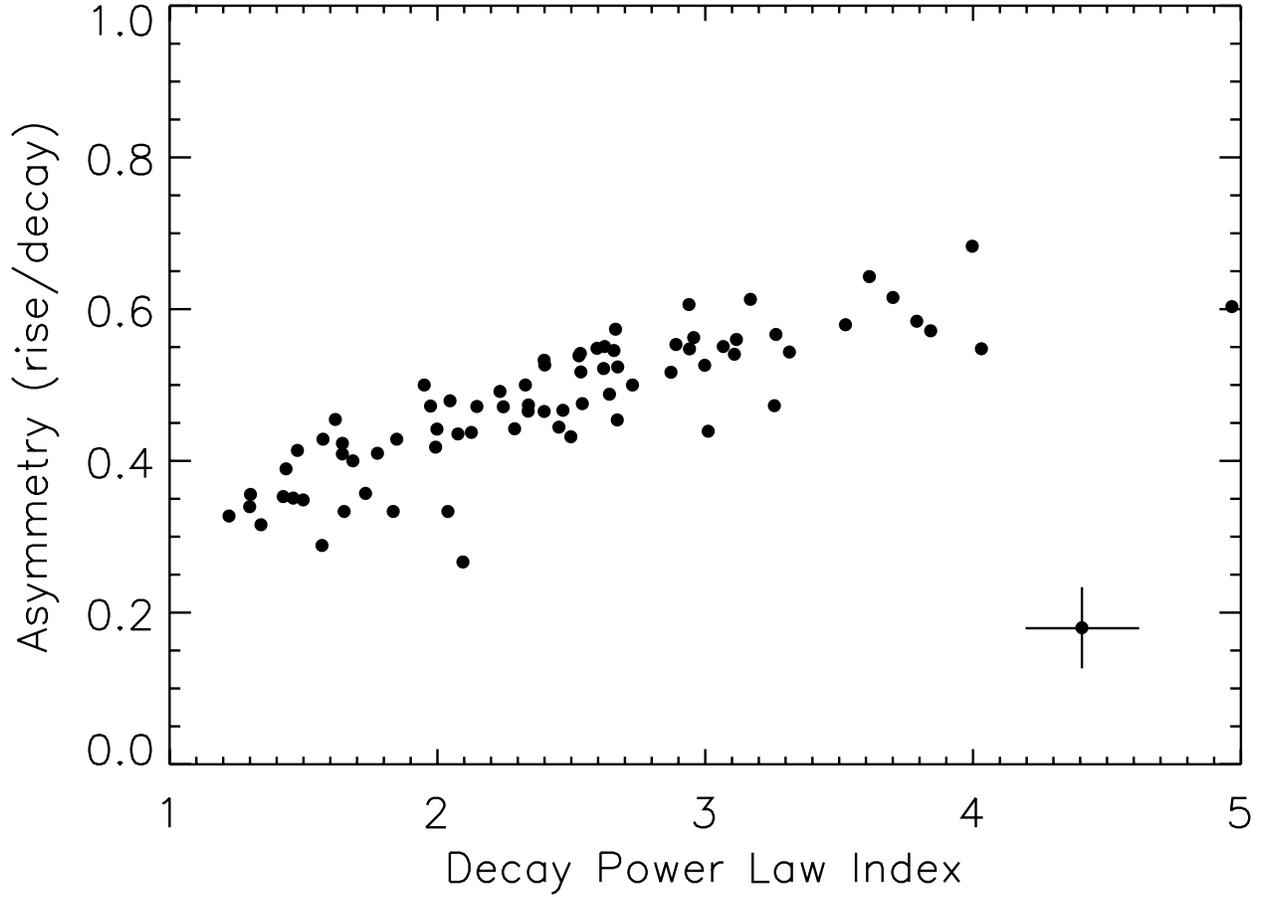}
\caption{Asymmetry plotted vs. the decay power-law index. Pulses
with higher decay indices appeared slightly less symmetric, with a
correlation strength of $R = 0.831$.  The average error is
illustrated in the bottom right corner.}
\end{figure}

\clearpage

\begin{figure} \label{fig:doppler}
\epsscale{1.0}
 \plotone{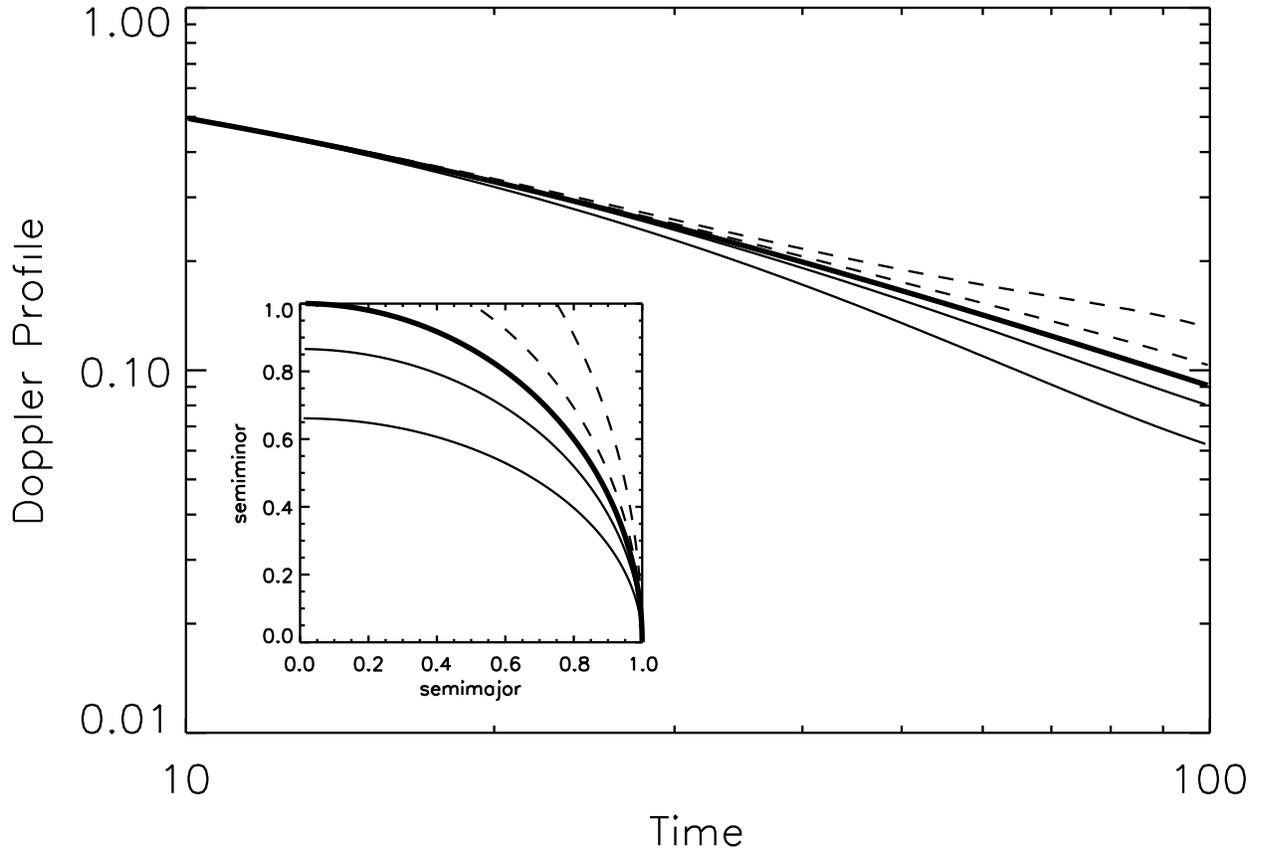}
\caption{Resulting Doppler profiles for various shell geometries
(shown in the inset).  The prolate shells (dashed) yield a Doppler
profile that is steeper than the spherical case (thick solid),
whereas the emission from oblate geometries (dashed) will be
observed to decay more slowly.}
\end{figure}

\clearpage

\begin{deluxetable}{rrrrrrrrr}
\label{tbl:data} \tablecolumns{7} \tablewidth{0pc}
\tablecaption{Burst Sample with Select Model Parameters. }
\tablehead{ \colhead{Burst} & \colhead{Trigger} &
\colhead{$T_{max}$ \tablenotemark{1}}  & \colhead{Peak Intensity}
& \colhead{$T_{90}$ \tablenotemark{2}} & \multicolumn{2}{c}{Power
Law Index}\tablenotemark{3} & \multicolumn{2}{c}{Model Fits
\tablenotemark{4}}
\\
\colhead{Name} & \colhead{Number} & \colhead{(s)} &
\colhead{(photons s$^{-1}$ cm$^{-2}$)} & \colhead{(s)} &
\colhead{Rise} & \colhead{Decay} & \colhead{$\chi_{p}^{2}$} &
\colhead{$\chi_{e}^{2}$} }

\startdata

910721  &   563 &   1.65    &   2.13    $\pm$   0.29    &   24.8   $\pm$   3.97    &   0.77    $\pm$   0.15    &   1.34    $\pm$   0.07    &   1.34    &   1.34    \\
911016  &   907 &   1.47    &   3.74    $\pm$   0.34    &   19.7   $\pm$   2.56    &   0.85    $\pm$   0.11    &   1.46    $\pm$   0.05    &   1.29    &   1.29    \\
911022  &   914 &   0.55    &   2.64    $\pm$   0.33    &   6.27    $\pm$   0.76    &   1.81    $\pm$   0.10    &   2.33    $\pm$   0.14    &   0.86    &   0.96    \\
911031  &   973 &   2.87    &   5.71    $\pm$   0.41    &   17.9   $\pm$   0.99    &   1.23    $\pm$   0.08    &   3.07    $\pm$   0.10    &   1.34    &   1.35    \\
911031  &   973 &   23.9    &   3.17    $\pm$   0.40    &   17.9   $\pm$   2.03    &   1.21    $\pm$   0.23    &   2.15    $\pm$   0.09    &   1.27    &   1.33    \\
911104  &   999 &   3.94    &   12.4   $\pm$   0.65    &   1.98    $\pm$   0.20    &   2.40    $\pm$   0.33    &   2.73    $\pm$   0.07    &   1.24    &   3.05    \\
911209  &   1157    &   4.83    &   12.2   $\pm$   0.55    &   8.58    $\pm$   1.07    &   1.60    $\pm$   0.58    &   1.62    $\pm$   0.10    &   1.30    &   1.58    \\
920216  &   1406    &   3.14    &   2.10    $\pm$   0.27    &   25.1   $\pm$   3.18    &   0.95    $\pm$   0.07    &   2.29    $\pm$   0.11    &   1.06    &   1.07    \\
920307  &   1467    &   4.32    &   2.59    $\pm$   0.27    &   18.3    $\pm$   2.05    &   2.53    $\pm$   0.08    &   2.62    $\pm$   0.08    &   1.01    &   1.06    \\
920801  &   1733    &   3.30     &   3.20    $\pm$   0.32    &   17.3   $\pm$   1.73    &   1.97    $\pm$   0.08    &   2.05    $\pm$   0.06    &   1.17    &   1.29    \\
920830  &   1883    &   1.26    &   5.36    $\pm$   0.37    &   8.19    $\pm$   0.51    &   1.48    $\pm$   0.04    &   2.60    $\pm$   0.08    &   1.06    &   1.07    \\
920925  &   1956    &   2.81    &   2.80    $\pm$   0.28    &   10.7   $\pm$   1.04    &   1.31    $\pm$   0.38    &   3.31    $\pm$   0.22    &   1.47    &   1.50    \\
921015  &   1989    &   116     &   2.96    $\pm$   0.31    &   18.3    $\pm$   1.83    &   2.00    $\pm$   0.18    &   2.40    $\pm$   0.06    &   1.35    &   1.66    \\
921207  &   2083    &   8.59    &   46.6   $\pm$   0.92     &   4.00       $\pm$   0.20    &   2.12    $\pm$   0.05    &   1.85    $\pm$   0.02    &   1.25    &   1.71    \\
921218  &   2102    &   1.88    &   1.79    $\pm$   0.26    &   10.9   $\pm$   1.51    &   1.90    $\pm$   1.03    &   2.94    $\pm$   0.29    &   0.89    &   0.95    \\
930120  &   2138    &   1.50     &   7.08    $\pm$   0.39    &   9.73    $\pm$   1.27    &   2.00    $\pm$   0.06    &   4.00    $\pm$   0.20    &   1.28    &   1.15    \\
930120  &   2138    &   48.9    &   35.6   $\pm$   0.59     &   9.73    $\pm$   1.77    &   1.46    $\pm$   0.47    &   2.23    $\pm$   0.29    &   0.96    &   1.00    \\
930120  &   2138    &   86.1    &   9.81    $\pm$   0.10    &   15.0      $\pm$   1.18    &   0.42    $\pm$   0.06    &   2.10    $\pm$   0.11    &   1.29    &   1.35    \\
930214  &   2193    &   10.9    &   1.78    $\pm$   0.26    &   70.3   $\pm$   10.3   &   0.80    $\pm$   0.01    &   2.50    $\pm$   0.04    &   1.04    &   1.02    \\
930612  &   2387    &   6.57    &   4.11    $\pm$   0.33    &   32.8   $\pm$   2.91    &   1.20    $\pm$   0.04    &   2.87    $\pm$   0.06    &   1.07    &   1.17    \\
930807  &   2484    &   1.87    &   1.75    $\pm$   0.27    &   13.5    $\pm$   1.87    &   1.26    $\pm$   0.29    &   3.26    $\pm$   0.43    &   1.10    &   1.17    \\
930909  &   2519    &   63.5    &   1.68    $\pm$   0.28    &   15.0   $\pm$   3.21    &   1.69    $\pm$   0.16    &   1.43    $\pm$   0.12    &   1.13    &   1.21    \\
930914  &   2530    &   114     &   2.03    $\pm$   0.27    &   25.3  $\pm$   3.28    &   1.12    $\pm$   0.05    &   2.34    $\pm$   0.12    &   1.13    &   1.14    \\
931127  &   2662    &   1.10     &   1.82    $\pm$   0.29    &   14.9   $\pm$   2.94    &   1.20    $\pm$   0.05    &   1.30    $\pm$   0.03    &   1.07    &   1.04    \\
931128  &   2665    &   1.34    &   2.10    $\pm$   0.31    &   12.8    $\pm$   2.25    &   1.74    $\pm$   0.43    &   1.65    $\pm$   0.10    &   1.00    &   1.16    \\
931221  &   2700    &   53.8    &   4.17    $\pm$   0.35    &   13.0   $\pm$   1.13    &   0.75    $\pm$   0.09    &   3.01    $\pm$   0.18    &   1.33    &   1.52    \\
940313  &   2880    &   0.4     &   3.20    $\pm$   0.29    &   4.16    $\pm$   0.54    &   1.32    $\pm$   0.27    &   2.13    $\pm$   0.15    &   1.51    &   1.61    \\
940410  &   2919    &   0.30     &   6.11    $\pm$   0.39    &   10.6   $\pm$   0.80    &   1.55    $\pm$   0.14    &   1.97    $\pm$   0.04    &   1.28    &   1.55    \\
940529  &   3003    &   9.56    &   3.04    $\pm$   0.32    &   28.8    $\pm$   3.13    &   1.51    $\pm$   0.04    &   2.62    $\pm$   0.08    &   1.11    &   1.21    \\
940830  &   3143    &   0.63    &   2.69    $\pm$   0.29    &   5.5     $\pm$   0.76    &   1.84    $\pm$   0.68    &   2.40    $\pm$   0.14    &   1.04    &   1.16    \\
940904  &   3155    &   0.66    &   1.92    $\pm$   0.26    &   3.39    $\pm$   0.46    &   1.24    $\pm$   0.21    &   3.84    $\pm$   0.60    &   0.88    &   1.01    \\
941023  &   3256    &   1.25    &   1.89    $\pm$   0.22    &   14.0   $\pm$   2.68    &   0.80    $\pm$   0.02    &   1.50    $\pm$   0.04    &   1.28    &   1.16    \\
941026  &   3257    &   3.27    &   3.21    $\pm$   0.27    &   36.8    $\pm$   5.49    &   0.52    $\pm$   0.04    &   1.57    $\pm$   0.07    &   1.02    &   1.08    \\
941026  &   3259    &   4.65    &   2.28    $\pm$   0.23    &   25.0      $\pm$   3.28    &   1.55    $\pm$   0.17    &   2.89    $\pm$   0.16    &   0.96    &   1.15    \\
941121  &   3290    &   2.91    &   10.9   $\pm$   0.36     &   1.98    $\pm$   0.47    &   1.47    $\pm$   0.25    &   2.04    $\pm$   0.09    &   0.92    &   1.52    \\
950211  &   3415    &   0.25    &   10.3   $\pm$   0.40     &   6.91    $\pm$   1.77    &   2.76    $\pm$   2.14    &   1.68    $\pm$   0.15    &   0.99    &   1.22    \\
950211  &   3415    &   11.5    &   2.36    $\pm$   0.11    &   3.78    $\pm$   0.34    &   1.26    $\pm$   0.14    &   2.47    $\pm$   0.17    &   1.48    &   1.36    \\
950624  &   3648    &   2.65    &   5.91    $\pm$   0.31    &   22.0      $\pm$   3.97    &   0.84    $\pm$   0.44    &   2.67    $\pm$   0.60    &   0.82    &   0.82    \\
950624  &   3648    &   24.1    &   36.6   $\pm$   1.22     &   17.0      $\pm$   3.51    &   1.10    $\pm$   0.04    &   2.40    $\pm$   0.14    &   1.03    &   0.97    \\
950624  &   3648    &   40.9    &   1243    $\pm$   3.71    &   11.7   $\pm$   1.16    &   2.54    $\pm$   0.07    &   1.57    $\pm$   0.03    &   1.45    &   1.19    \\
950818  &   3765    &   66.1    &   25.9   $\pm$   0.54     &   8.83    $\pm$   0.98    &   2.42    $\pm$   0.07    &   1.84    $\pm$   0.02    &   1.50    &   1.43    \\
951016  &   3870    &   0.43    &   14.1   $\pm$   0.46     &   5.76    $\pm$   0.52    &   1.38    $\pm$   0.09    &   1.42    $\pm$   0.02    &   1.52    &   1.78    \\
951019  &   3875    &   0.21    &   3.02    $\pm$   0.23    &   4.03    $\pm$   0.42    &   0.96    $\pm$   0.11    &   1.73    $\pm$   0.13    &   1.57    &   1.51    \\
951030  &   3886    &   0.20     &   2.27    $\pm$   0.21    &   2.82    $\pm$   0.65    &   1.27    $\pm$   0.12    &   2.45    $\pm$   0.27    &   1.00    &   1.55    \\
951102  &   3892    &   0.61    &   1.86    $\pm$   0.20    &   5.31    $\pm$   0.51    &   1.26    $\pm$   0.24    &   2.67    $\pm$   0.29    &   0.97    &   1.03    \\
951213  &   3954    &   0.70     &   8.45    $\pm$   0.39    &   10.3    $\pm$   0.70    &   4.31    $\pm$   3.52    &   2.54    $\pm$   0.50    &   1.06    &   1.52    \\
951228  &   4157    &   7.67    &   2.32    $\pm$   0.24    &   17.7   $\pm$   2.16    &   1.92    $\pm$   0.09    &   2.53    $\pm$   0.12    &   1.21    &   1.36    \\
960113  &   4350    &   0.51    &   3.52    $\pm$   0.25    &   6.66    $\pm$   0.83    &   2.41    $\pm$   0.16    &   1.95    $\pm$   0.07    &   1.29    &   1.86    \\
960113  &   4350    &   14.3    &   1.65    $\pm$   0.27    &   6.66    $\pm$   0.75    &   1.00    $\pm$   0.02    &   2.00    $\pm$   0.03    &   1.39    &   1.22    \\
960113  &   4350    &   34.1    &   8.71    $\pm$   0.45    &   6.66    $\pm$   0.98    &   1.70    $\pm$   0.34    &   2.66    $\pm$   0.23    &   1.30    &   1.47    \\
960114  &   4368    &   2.45    &   58.6    $\pm$   0.83    &   10.0      $\pm$   0.75    &   2.87    $\pm$   0.68    &   2.53    $\pm$   0.07    &   1.26    &   2.01    \\
960530  &   5478    &   1.98    &   3.10    $\pm$   0.24    &   14.3   $\pm$   1.43    &   0.98    $\pm$   0.08    &   2.54    $\pm$   0.17    &   0.97    &   0.98    \\
960530  &   5478    &   261     &   7.69    $\pm$   1.70    &   37.2   $\pm$   6.22    &   1.20    $\pm$   0.02    &   3.00    $\pm$   0.08    &   1.12    &   1.16    \\
960613  &   5495    &   0.23    &   2.39    $\pm$   0.23    &   5.44    $\pm$   1.66    &   0.63    $\pm$   0.34    &   1.65    $\pm$   0.16    &   1.16    &   1.14    \\
960624  &   5517    &   0.83    &   2.06    $\pm$   0.23    &   7.42    $\pm$   0.79    &   1.72    $\pm$   0.11    &   3.61    $\pm$   0.37    &   1.04    &   1.05    \\
960628  &   5523    &   1.03    &   3.73    $\pm$   0.28    &   6.98    $\pm$   0.71    &   1.92    $\pm$   0.07    &   3.17    $\pm$   0.15    &   1.19    &   1.34    \\
960715  &   5541    &   1.32    &   1.84    $\pm$   0.23    &   12.6    $\pm$   1.68    &   1.51    $\pm$   0.47    &   2.96    $\pm$   0.39    &   1.13    &   1.16    \\
960912  &   5601    &   7.55    &   4.94    $\pm$   0.29    &   9.00    $\pm$   0.74    &   1.30    $\pm$   0.26    &   2.64    $\pm$   0.16    &   1.29    &   1.38    \\
970405  &   6159    &   3.20     &   2.16    $\pm$   0.25    &   12.0    $\pm$   1.79    &   1.82    $\pm$   0.13    &   1.30    $\pm$   0.11    &   1.18    &   1.28    \\
970815  &   6335    &   98.3    &   4.29    $\pm$   0.29    &   21.3   $\pm$   3.02    &   1.29    $\pm$   0.08    &   3.52    $\pm$   0.40    &   1.10    &   1.17    \\
970925  &   6397    &   3.30     &   6.02    $\pm$   0.30    &   20.4   $\pm$   1.44    &   1.25    $\pm$   0.07    &   2.25    $\pm$   0.05    &   1.61    &   1.53    \\
971127  &   6504    &   3.12    &   2.79    $\pm$   0.24    &   22.1   $\pm$   3.00    &   1.18    $\pm$   0.12    &   1.78    $\pm$   0.08    &   1.00    &   1.09    \\
980301  &   6621    &   32.5    &   6.85    $\pm$   0.34    &   7.36    $\pm$   0.47    &   1.51    $\pm$   0.05    &   1.65    $\pm$   0.03    &   1.21    &   1.59    \\
980302  &   6625    &   5.12    &   2.04    $\pm$   0.26    &   31.4   $\pm$   4.19    &   1.32    $\pm$   0.05    &   3.12    $\pm$   0.19    &   1.04    &   1.16    \\
980401  &   6672    &   6.85    &   5.60    $\pm$   0.29    &   3.80     $\pm$   0.41    &   1.17    $\pm$   0.07    &   2.34    $\pm$   0.17    &   0.78    &   0.84    \\
980718  &   6930    &   31.7    &   5.76    $\pm$   0.36    &   6.34    $\pm$   0.30    &   1.51    $\pm$   0.04    &   3.70    $\pm$   0.16    &   1.15    &   1.10    \\
990102  &   7293    &   3.53    &   3.06    $\pm$   0.23    &   25.2   $\pm$   2.92    &   0.96    $\pm$   0.07    &   1.99    $\pm$   0.07    &   1.20    &   1.20    \\
990102  &   7295    &   2.14    &   3.50    $\pm$   0.35    &   8.50     $\pm$   1.05    &   1.20    $\pm$   0.42    &   2.08    $\pm$   0.35    &   0.97    &   0.98    \\
990316  &   7475    &   8.54    &   3.87    $\pm$   0.29    &   34.4   $\pm$   2.96    &   0.99    $\pm$   0.07    &   4.03    $\pm$   0.19    &   1.12    &   1.05    \\
990505  &   7548    &   3.7     &   3.14    $\pm$   0.25    &   10.6   $\pm$   1.02    &   1.92    $\pm$   0.46    &   2.94    $\pm$   0.17    &   0.82    &   1.04    \\
990528  &   7588    &   2.76    &   2.32    $\pm$   0.24    &   15.5   $\pm$   1.77    &   1.15    $\pm$   0.13    &   3.11    $\pm$   0.25    &   0.83    &   0.92    \\
990707  &   7638    &   1.12    &   1.92    $\pm$   0.22    &   15.0   $\pm$   2.68    &   1.27    $\pm$   0.30    &   1.22    $\pm$   0.06    &   1.30    &   1.33    \\
990712  &   7648    &   4.45    &   1.78    $\pm$   0.21    &   24.7    $\pm$   3.45    &   1.22    $\pm$   0.13    &   3.79    $\pm$   0.41    &   1.05    &   1.17    \\
990816  &   7711    &   1.86    &   3.97    $\pm$   0.26    &   13.2   $\pm$   1.32    &   0.85    $\pm$   0.04    &   3.26    $\pm$   0.24    &   1.34    &   1.48    \\
323 &   8049        &   30.5    &   1.77    $\pm$   0.19    &   36.6   $\pm$   4.23    &   1.13    $\pm$   0.04    &   4.97    $\pm$   0.42    &   1.04    &   1.24    \\
519 &   8111        &   4.98    &   4.33    $\pm$   0.28    &   13.8   $\pm$   1.68    &   1.69    $\pm$   0.03    &   1.48    $\pm$   0.02    &   1.36    &   1.27    \\

\tablenotetext{1}{Time of peak emission.}
\tablenotetext{2}{Interval between times when 5$\%$ and 95$\%$ of
burst counts (above background) accumulate for each individual
pulse.} \tablenotetext{3}{As measured by the power law rise model
(Eq. \ref{F1}).} \tablenotetext{4}{The $\chi_{2}$ values for the
power law (Eq. \ref{F1}) and exponential rise (Eq. \ref{F2})
models respectively.}

\enddata
\end{deluxetable}

\end{document}